\RequirePackage{ifpdf}
\ifpdf % We are running pdfTeX in pdf mode
\documentclass[pdftex]{sigma}
\else
\documentclass{sigma}
\fi

\numberwithin{equation}{section}

\begin{document}

\allowdisplaybreaks

\renewcommand{\PaperNumber}{020}

\renewcommand{\thefootnote}{$\star$}

\FirstPageHeading

\ShortArticleName{Clif\/ford Algebra Derivations of Tau-Functions}

\ArticleName{Clif\/ford Algebra Derivations of Tau-Functions\\
for Two-Dimensional Integrable Models\\
with Positive and Negative Flows\footnote{This paper is a
contribution to the Vadim Kuznetsov Memorial Issue ``Integrable
Systems and Related Topics''. The full collection is available at
\href{http://www.emis.de/journals/SIGMA/kuznetsov.html}{http://www.emis.de/journals/SIGMA/kuznetsov.html}}}

\Author{Henrik ARATYN~$^\dag$ and Johan VAN DE LEUR~$^\ddag$}

\AuthorNameForHeading{H. Aratyn and J. van de Leur}

\Address{$^\dag$~Department of Physics,
University of Illinois at Chicago,\\
$\phantom{^{\dag}}$~845 W. Taylor St., Chicago, IL 60607-7059,
USA} \EmailD{\href{mailto:aratyn@uic.edu}{aratyn@uic.edu}}

\Address{$^\ddag$~Mathematical Institute, University of Utrecht, \\
$\phantom{^{\ddag}}$~P.O. Box 80010, 3508 TA Utrecht, The
Netherlands}
\EmailD{\href{mailto:vdleur@math.uu.nl}{vdleur@math.uu.nl}}

\ArticleDates{Received October 11, 2006, in f\/inal form January
09, 2007; Published online February 06, 2007}

\Abstract{We use a Grassmannian framework to def\/ine
multi-component tau functions as expectation values of certain
multi-component Fermi operators satisfying  simple bilinear
commutation relations on Clif\/ford algebra. The tau functions
contain both positive and negative f\/lows and are shown to
satisfy the $2n$-component KP hierarchy. The hierarchy equations
can  be formulated in terms of pseudo-dif\/ferential equations for
$n \times n$ matrix wave functions derived in terms of tau
functions. These equations are cast in form of Sato--Wilson
relations. A reduction process leads to the AKNS, two-component
Camassa--Holm and Cecotti--Vafa models and the formalism provides
simple formulas for their solutions.}


\Classification{11E88; 17B67; 22E67; 37K10}

\begin{quote}
{\bfseries\itshape In memory of Vadim Kuznetsov.} \it One of us
(JvdL) f\/irst met Vadim at a~semi\-nar on Quantum Groups held at
the Korteweg--de Vries Institute in Amsterdam in 1993. Vadim was
then still a post-Doc. Later meetings at several other
conferences, e.g. in Cambridge, Oberwolfach and Montreal provided
better opportunities to learn Vadim's kind personality and his
wonderful sense of humor. Both authors of this paper have the best
recollection of Vadim from meeting at the conference ``Classical
and Quantum Integrable Systems'' organized in 1998 in Oberwolfach
by Werner Nahm and Pierre van Moerbeke. A memory of one pleasant
evening spent with him in the setting of a beautiful conference
center library clearly stands out. Vadim joined us and Boris
Konopelchenko after a (successful) play of pool game against Boris
Dubrovin. The library has a wonderful wine cellar and we were all
having a good time. Vadim was in great mood and filled the
conversation with jokes and funny anecdotes. When thinking of him,
we will always remember in particular that most enjoyable evening.
\end{quote}

\renewcommand{\thefootnote}{\arabic{footnote}}
\setcounter{footnote}{0}

\section{Introduction}
Tau-functions are the building blocks of integrable models. The
Grassmannian techniques have been shown in the past to be very
ef\/fective in theory of  tau-functions. In this paper we exploit
a Grassmannian approach to constructing tau functions in terms of
expectation values of certain Fermi operators constructed using
boson-fermion correspondence. This formalism provides a systematic
way of constructing multi-component tau functions for
multi-dimensional Toda models, which, in this paper, embed both
positive and negative f\/lows.

The set of Hirota equations for the tau functions is obtained by
taking expectation value of both sides of bilinear commutation
relation
\begin{gather*}
 A \otimes A \Omega = \Omega  A \otimes A,
 \qquad
 \Omega = \sum_{{\mathfrak l},i} \psi^{+\,(i)}_{\mathfrak l} \otimes
 \psi^{-\,(i)}_{-{\mathfrak l}}
 \end{gather*}
def\/ined on a Clif\/ford algebra with Fermi operators $\psi^{\pm
\,(i)}_{\mathfrak l}$, satisfying the relations
\begin{gather*}
\psi^{\lambda \,(i)}_{\mathfrak l}\psi^{\mu\,(j)}_{\mathfrak k}+
\psi^{\mu\,(j)}_{\mathfrak k}\psi^{\lambda \,(i)}_{\mathfrak l}=
\delta_{i,j}\delta_{\lambda,\mu}\delta_{{\mathfrak l},-{\mathfrak
k}},
\end{gather*}
for ${\mathfrak l,k} \in \mathbb{Z}+1/2$, $i,j=1,\dots ,n$,
$\lambda,\mu=+,-$.

We rewrite Hirota equations in terms of formal
pseudo-dif\/ferential operators acting on matrix wave functions
derived from the tau functions. This method gives rise to a
general set of Sato--Wilson equations.

Next, we impose a set conditions on the tau function which
def\/ine  a reduction process.  Under this reduction process the
pseudo-dif\/ferential equations for the wave functions describe
f\/lows of dressing matrices of the multi-dimensional Toda model.
In the case of $2\times 2$ matrix wave functions these equations
embed the AKNS model and the two-component version of the
Camassa--Holm (CH) model for, respectively, positive and negative
f\/lows of the multi-dimensional $2 \times 2$ Toda model.

Section 2 is meant as an informal review of semi-inf\/inite wedge
space and Clif\/ford algebra. In this setup, in Section 3, we
formulate the multi-component tau functions as expectation values
of operators satisfying the bilinear identity. This formalism
contains the Toda lattice hierarchy as a special case. Section 4
shows how to rewrite the formalism in terms of
pseudo-dif\/ferential operators acting an wave function. We arrive
in this way in general equations of Sato--Wilson type. The
objective of the next Section 5 is to introduce a general
reduction process leading to a multi-dimensional Toda model with
positive and negative f\/lows acting on $n \times n$ matrix wave
functions explicitly found in terms of components of the tau
functions from Section 3. The Grassmannian method provides in this
section explicit construction of matrices solving the
Riemann--Hilbert factorization problem for $GL_n$.

As shown in Section 6 the model obtained in Section 5 embeds both
AKNS and the 2-com\-ponent Camassa--Holm equations. Further
reduction uses an automorphism of order $4$ and as described in
Section 7 reduces f\/low equations to Cecotti--Vafa equations. In
Section 8, we use the Virasoro algebra constraint to further
reduce the model by imposing homogeneity relations on matrices
satisfying Cecotti--Vafa equations.

The particular advantage of our construction is that it leads to
solutions of the AKNS and the 2-component Camassa--Holm equations
and Cecotti--Vafa equations in terms of relatively simple
correlation functions involving Fermi operators def\/ined
according to the Fermi--Bose correspondence. These solutions are
constructed in Section 9 and 10, respectively.

\section{Semi-inf\/inite wedge space and Clif\/ford algebra}

Following \cite{KL}, we introduce the semi-inf\/inite wedge space
$F =\Lambda^{\frac{1}{2}\infty} {\mathbb C}^{\infty}$ as the
vector space with a basis consisting of all semi-inf\/inite
monomials of the form $v_{{\mathfrak i}_{1}} \wedge v_{{\mathfrak
i}_{2}} \wedge v_{{\mathfrak i}_{3}} \cdots$, with ${\mathfrak
i}_j\in{1}/{2}+\mathbb{Z}$, where ${\mathfrak i}_{1} > {\mathfrak
i}_{2} > {\mathfrak i}_{3} > \cdots$ and ${\mathfrak i}_{\ell +1}
= {\mathfrak i}_{\ell} -1$ for $\ell \gg 0$.  Def\/ine the wedging
and contracting operators $\psi^{+}_{\mathfrak j}$ and
$\psi^{-}_{\mathfrak j}\ \ ({\mathfrak j} \in {\mathbb
Z}+{1}/{2})$ on $F$ by
\begin{gather*}
\psi^{+}_{\mathfrak j} (v_{{\mathfrak i}_{1}} \wedge v_{{\mathfrak
i}_{2}} \wedge \cdots ) = \begin{cases} 0 & \text{if}\ -{\mathfrak
j} = {\mathfrak i}_{s}\ \text{for some}\ s,
\\
(-1)^{s} v_{{\mathfrak i}_{1}} \wedge v_{{\mathfrak i}_{2}} \cdots
\wedge v_{{\mathfrak i}_{s}} \wedge v_{-{\mathfrak j}} \wedge
v_{{\mathfrak i}_{s+1}} \wedge \cdots &\text{if}\ {\mathfrak
i}_{s} > -{\mathfrak j} > {\mathfrak i}_{s+1},\end{cases}
\\
\psi^{-}_{\mathfrak j} (v_{{\mathfrak i}_{1}} \wedge v_{{\mathfrak
i}_{2}} \wedge \cdots ) = \begin{cases} 0
&\text{if}\ {\mathfrak j} \neq {\mathfrak i}_{s}\ \text{for all}\ s, \\
(-1)^{s+1} v_{{\mathfrak i}_{1}} \wedge v_{{\mathfrak i}_{2}}
\wedge \cdots \wedge v_{{\mathfrak i}_{s-1}} \wedge v_{{\mathfrak
i}_{s+1}} \wedge \cdots &\text{if}\ {\mathfrak j} = {\mathfrak
i}_{s}.
\end{cases}
\end{gather*}
These operators satisfy the following relations (${\mathfrak
i},{\mathfrak j} \in {\mathbb Z}+{1}/{2}$, $\lambda,\mu =+,-$):
\begin{gather}
\psi^{\lambda}_{\mathfrak i} \psi^{\mu}_{\mathfrak j} +
\psi^{\mu}_{\mathfrak j} \psi^{\lambda}_{\mathfrak i} =
\delta_{\lambda,-\mu} \delta_{{\mathfrak i},-{\mathfrak j}},
\label{comm}
\end{gather}
hence they generate a Clif\/ford algebra, which we denote by
${\cal C}\ell$.

Introduce the following elements of $F$ $(m \in {\mathbb Z})$:
\begin{gather*}
|m\rangle = v_{m-\frac{1}{2} } \wedge v_{m-\frac{3}{2} } \wedge
v_{m-\frac{5}{2} } \wedge \cdots.\end{gather*} It is clear that
$F$ is an irreducible ${\cal C}\ell$-module such that
\begin{gather*}
\psi^{\pm}_{\mathfrak j} |0\rangle = 0 \quad \text{for}\quad
{\mathfrak j} > 0.
\end{gather*}
Think of the adjoint module $F^*$ in the following way, it is the
vector space with a basis consisting of all semi-inf\/inite
monomials of the form $\cdots \wedge v_{{\mathfrak i}_{3}} \wedge
v_{{\mathfrak i}_{2}} \wedge v_{{\mathfrak i}_{1}} $, where
${\mathfrak i}_{1} < {\mathfrak i}_{2} < {\mathfrak i}_{3} <
\cdots$ and ${\mathfrak i}_{\ell +1} = {\mathfrak i}_{\ell} +1$
for $\ell \gg 0$. The operators $\psi^{+}_{\mathfrak j}$ and
$\psi^{-}_{\mathfrak j}$ $({\mathfrak j} \in {\mathbb Z}+{1}/{2})$
also act on $F^*$ by contracting and wedging, but in a dif\/ferent
way, viz.,
\begin{gather*}
(\cdots\wedge v_{{\mathfrak i}_{2}} \wedge v_{{\mathfrak i}_{1}}
)\psi^{+}_{\mathfrak j}  = \begin{cases} 0
& \text{if}\ {\mathfrak j}\ne {\mathfrak i}_{s}\ \text{for all}\ s, \\
(-1)^{s+1} \cdots\wedge v_{{\mathfrak i}_{s+1}}\wedge
v_{{\mathfrak i}_{s-1}} \wedge \cdots v_{{\mathfrak i}_{2}} \wedge
v_{{\mathfrak i}_{1}}&\text{if}\ {\mathfrak i}_{s} = {\mathfrak
j},\end{cases}
\\[1ex]
(\cdots\wedge v_{{\mathfrak i}_{2}} \wedge v_{{\mathfrak i}_{1}}
)\psi^{-}_{\mathfrak j} = \begin{cases} 0 &\text{if}
\  -{\mathfrak j} = {\mathfrak i}_{s}\ \text{for some}\ s, \\
(-1)^{s} \cdots\wedge v_{{\mathfrak i}_{s+1}}\wedge v_{\mathfrak
j}\wedge v_{{\mathfrak i}_{s}} \wedge \cdots v_{{\mathfrak i}_{2}}
\wedge v_{{\mathfrak i}_{1}}&\text{if}\ {\mathfrak
i}_{s}<-{\mathfrak j}<{\mathfrak i}_{s+1}.
\end{cases}
\end{gather*}
We introduce the element $\langle m |$ by $\langle m | = \cdots
\wedge v_{m+\frac{5}{2} } \wedge v_{m+\frac{3}{2} } \wedge
v_{m+\frac{1}{2}},$ such that $\langle 0 |\psi^{\pm}_{\mathfrak j}
= 0$ for ${\mathfrak j} < 0$. We def\/ine the vacuum expectation
value by $\langle 0 |0\rangle=1$, and denote $\langle
A\rangle=\langle 0 |A|0\rangle$.

Note that $(\psi^\pm_{\mathfrak k})^*=\psi_{-{\mathfrak k}}^{\mp}$
and that
\begin{gather}
|V({\mathfrak i}_1,\ldots, {\mathfrak i}_k,{\mathfrak j}_1,\ldots,
{\mathfrak j}_\ell)\rangle= \psi^+_{{\mathfrak
i}_1}\psi^+_{{\mathfrak i}_2}\cdots\psi^+_{{\mathfrak i}_k}
\psi^-_{{\mathfrak j}_1}\psi^-_{{\mathfrak
j}_2}\cdots\psi^-_{{\mathfrak j}_\ell}|0\rangle,\nonumber
\\
\langle V({\mathfrak i}_1,\ldots, {\mathfrak i}_k,{\mathfrak
j}_1,\ldots, {\mathfrak j}_\ell)| = \langle 0| \psi^+_{-{\mathfrak
j}_\ell}\psi^+_{-{\mathfrak j}_{\ell-1}}\cdots\psi^+_{-{\mathfrak
j}_1} \psi^-_{-{\mathfrak i}_k}\psi^-_{-{\mathfrak
i}_{k-1}}\cdots\psi^-_{-{\mathfrak i}_1}, \label{basis}
\end{gather}
with ${\mathfrak i}_1<{\mathfrak i}_2< \cdots < {\mathfrak i}_k<0$
and ${\mathfrak j}_1< {\mathfrak i}_2< \cdots< {\mathfrak
j}_\ell<0$ form dual basis of $F$ and $F^*$, i.e.,
\begin{gather}
\langle V({\mathfrak r}_1,\ldots, {\mathfrak r}_m,{\mathfrak
s}_1,\ldots,{\mathfrak s}_q) |V({\mathfrak i}_1,\ldots, {\mathfrak
i}_k,{\mathfrak j}_1,\ldots, {\mathfrak j}_\ell)\rangle
=\delta_{({\mathfrak r}_1,\ldots, {\mathfrak r}_m),({\mathfrak
i}_1,\ldots,{\mathfrak i}_k)} \delta_{({\mathfrak s}_1,\ldots,
{\mathfrak s}_q),({\mathfrak j}_1,\ldots, {\mathfrak j}_\ell)}.
\label{basis2}
\end{gather}
We relabel the basis vectors $v_{i}$ and with them the
corresponding fermionic operators (the wedging and contracting
operators). This relabeling can be done in many dif\/ferent ways,
see e.g.~\cite{TV}, the simplest one is the following
($j=1,2,\dots,n$):
\begin{gather*}
v^{({j})}_{\mathfrak k} = v_{n{\mathfrak k} -
\frac{1}{2}(n-2j+1)},
\end{gather*}
and correspondingly:
\begin{gather*}
\psi^{\pm (j)}_{\mathfrak k} = \psi^{\pm}_{n{\mathfrak k}
\pm\frac{1}{2}(n-2j+1)}.
\end{gather*}
Notice that with this relabeling we have:
\begin{gather*}
\psi^{\pm (j)}_{\mathfrak k}|0\rangle = 0\quad \text{for}\quad
{\mathfrak k} > 0.
\end{gather*}

Def\/ine {\it partial charges}  and {\it partial energy} by
\begin{gather*}
\text{charge}_{j}\ \psi^{\pm (i)}_{\mathfrak k}=\pm
\delta_{ij},\qquad \text{charge}_{j}\  |0\rangle =0,
\\
\text{energy}_{j}\ \psi^{\pm (i)}_{\mathfrak
k}=-\delta_{ij}k,\qquad \text{energy}_{j}\  |0\rangle =0.
\end{gather*}
Total charge and energy is def\/ined as the sum of partial
charges, respectively the sum of partial energy.

Introduce the  fermionic f\/ields $(0\ne z \in {\mathbb C})$:
\begin{gather*}
\psi^{\pm (j)}(z)=\sum_{{\mathfrak k} \in {\mathbb Z}+{1}/{2}}
\psi^{\pm (j)}_{\mathfrak k} z^{-{\mathfrak k}-\frac{1}{2}}.
\end{gather*}
Next, we introduce bosonic f\/ields ($1 \leq i,j \leq n$):
\begin{gather*}
\alpha^{(ij)}(z) \equiv \sum_{k \in {\mathbb Z}} \alpha^{(ij)}_{
k} z^{-{k}-1} = :\psi^{+(i)}(z) \psi^{-(j)}(z):,
\end{gather*}
where $:\ :$ stands for the {\it normal ordered product} def\/ined
in the usual way $(\lambda,\mu = +$ or $-$):
\begin{gather*}
:\psi^{\lambda (i)}_{\mathfrak k} \psi^{\mu (j)}_{\mathfrak l}: =
\begin{cases} \psi^{
\lambda (i)}_{\mathfrak k}
\psi^{\mu (j)}_{\mathfrak l}\ &\text{if}\ {\mathfrak l} > 0, \\
-\psi^{\mu (j)}_{\mathfrak l} \psi^{\lambda (i)}_{\mathfrak k}
&\text{if}\ {\mathfrak l} < 0.\end{cases}
\end{gather*}
One checks (using e.g. the Wick formula) that the operators
$\alpha^{(ij)}_{k}$ satisfy the commutation relations of the
af\/f\/ine algebra $gl_{n}({\mathbb C})^{\wedge}$ with the central
charge $1$, i.e.:
\begin{gather*}
{}[\alpha^{(ij)}_{p},\alpha^{(k\ell)}_{q}] =
\delta_{jk}\alpha^{(i\ell )}_{p+q} - \delta_{i\ell}
\alpha^{(kj)}_{p+q} + p\delta_{i\ell} \delta_{jk}\delta_{p,-q},
\end{gather*}
and that
\begin{gather*}
\alpha^{(ij)}_{k}|m\rangle = 0 \quad \text{if}\quad k > 0 \quad
\text{or}\quad k = 0\ \text{and}\ i < j.
\end{gather*}
The operators $\alpha^{(i)}_{k} \equiv \alpha^{(ii)}_{k}$ satisfy
the canonical commutation relation of the associative oscillator
algebra,  which we denote by ${\mathfrak a}$:
\begin{gather*}
{}[\alpha^{(i)}_{k},\alpha^{(j)}_{\ell}] =
k\delta_{ij}\delta_{k,-\ell},
\end{gather*}
and one has
\begin{gather*}
\alpha^{(i)}_{k}|m\rangle = 0 \quad \text{for}\quad k > 0,\qquad
\langle m|\alpha^{(i)}_{k}= 0 \quad \text{for}\quad k < 0.
\end{gather*}

Note that $\alpha_0^{(j)}$ is the operator that counts the $j$-th
charge. The $j$-th energy is counted by the operator
\begin{gather*}
-\sum_{{\mathfrak k}\in \frac{1}{2}+\mathbb{Z}} {\mathfrak
k}:\psi_{\mathfrak k}^{+(j)}\psi_{-{\mathfrak k}}^{-(j)}:.
\end{gather*}
The complete energy is counted by the sum over all $j$ of such
operators.
 In (\ref{h1}) we will def\/ine  another operator $L_0$, which will also count
 the complete energy.
In order to express the fermionic f\/ields $\psi^{\pm (i)}(z)$ in
terms of the bosonic f\/ields $\alpha^{(ii)}(z)$, we need some
additional operators~$Q_{i}$, $i = 1,2,\ldots, n$, on $F$.  These
operators are uniquely def\/ined by the following conditions:
\begin{gather}
Q_{i}|0\rangle = \psi^{+(i)}_{-\frac{1}{2}} |0\rangle,\qquad
Q_{i}\psi^{\pm(j)}_{\mathfrak k} = (-1)^{\delta_{ij}+1}
\psi^{\pm(j)}_{{\mathfrak k}\mp \delta_{ij}}Q_{i}. \label{Q}
\end{gather}
They satisfy the following commutation relations:
\begin{gather*}
Q_{i}Q_{j} = -Q_{j}Q_{i}\quad \text{if}\quad i \neq j,\qquad
[\alpha^{(i)}_{k},Q_{j}] =\delta_{ij} \delta_{k0}Q_{j}.
\end{gather*}

We shall use below the following notation
\begin{gather*}
|k_{1},k_{2}, \ldots, k_n\rangle = Q^{k_{1}}_{1}
Q^{k_{2}}_{2}\cdots Q^{k_{n}}_{n}|0\rangle,\qquad \langle
k_{1},k_{2}, \ldots, k_n|=\langle 0|Q^{-k_{n}}_{n}\cdots
Q^{-k_{2}}_{2}Q^{-k_{1}}_{1},
 \end{gather*}
 such that
 \begin{gather*}
 \langle k_{1},k_{2}, \ldots,k_n|k_{1},k_{2}, \ldots, k_n\rangle=\langle
0|0\rangle=1 .\end{gather*} One easily checks the following
relations:
\begin{gather*}
{}[\alpha^{(i)}_{k},\psi^{\pm (j)}_{\mathfrak m}] = \pm
\delta_{ij} \psi^{\pm (j)}_{k+{\mathfrak m}}
\end{gather*}
and
\begin{gather*}
Q_i^{\pm 1}|k_{1},k_{2},\ldots, k_n\rangle
=(-)^{k_1+k_2+\cdots+k_{i-1}}|k_{1},k_{2},\ldots,k_{i-1},k_i\pm
1,k_{i+1}, \ldots,k_n\rangle,
\\
\langle k_{1},k_{2}, \ldots,k_n|Q_i^{\pm 1}=
(-)^{k_1+k_2+\cdots+k_{i-1}}\langle
k_{1},k_{2},\ldots,k_{i-1},k_i\mp 1,k_{i+1} ,\ldots, k_n|.
\end{gather*}
These formula's lead to the following vertex operator expression
for $\psi^{\pm (i)}(z)$. Given any sequence $s=(s_1,s_2,\dots)$,
def\/ine
\begin{gather*}
\Gamma^{(j)}_\pm( s)=\exp \left(\sum_{k=1}^\infty s_k
\alpha^{(j)}_{\pm k}\right),
\end{gather*}
then
\begin{theorem}[\cite{DJKM1,JM}]
\begin{gather*}
\psi^{\pm (i)}(z) = Q^{\pm 1}_{i}z^{\pm \alpha^{(i)}_{0}} \exp
\Biggl(\mp \sum_{k < 0} \frac{1}{k} \alpha^{(i)}_{k}z^{-k}\Biggr)
\exp\Biggl(\mp\sum_{k > 0} \frac{1}{k} \alpha^{(i)}_{k}
z^{-k}\Biggr)
\\
\phantom{\psi^{\pm (i)}(z)} {}=Q^{\pm 1}_{i}z^{\pm
\alpha^{(i)}_{0}}\Gamma_{-}^{(i)}(\pm [z])\Gamma_{+}^{(i)}(\mp
[z^{-1}]),
 \end{gather*}
 where $[z]=\big(z,\frac{z^2}{2}, \frac{z^3}{3},\ldots\big)$.
\end{theorem}

Note,
\begin{gather*}
\Gamma_+^{(j)}(s) \,|k_{1},k_{2}, \ldots,k_n\rangle =
|k_{1},k_{2}, \ldots,k_n\rangle,\nonumber \qquad \langle
k_{1},k_{2}, \ldots, k_n| \Gamma_-^{(j)}(s) = \langle k_{1},k_{2},
\ldots,k_n|. \label{Gamma_vac}
\end{gather*}
Also observe that $(\Gamma_{\pm}^{(j)})^*=\Gamma_{\mp}^{(j)}$ and
\begin{gather*}
\Gamma_+^{(j)}(s)\, \Gamma_-^{(k)}(s') =
\gamma(s,s')^{\delta{jk}}\, \Gamma_-^{(k)}(s')\,
\Gamma_+^{(j)}(s), \label{Gamma_Gamma}
\end{gather*}
where
\begin{gather*}
\gamma(s,s')=e^{\sum n\, s_n s'_n}.
\end{gather*}
We have
\begin{gather*}
\Gamma_\pm^{(j)}(s) \, \psi^{+(k)}(z)  = \gamma(s,[z^{\pm
1}])^{\delta_{jk}} \psi^{+(k)}(z) \, \Gamma_\pm^{(j)}(s), \notag
\\
\Gamma_\pm^{(j)}(s) \, \psi^{-(k)}(z)  = \gamma(s,-[z^{\pm
1}])^{\delta_{jk}} \psi^{-(k)}(z) \, \Gamma_\pm^{(j)} (s).
\label{e111}
\end{gather*}
Note that
\begin{gather*}
\gamma(t,[z])= \exp\Biggl(\sum_{n\ge 1} t_n\, z^n\Biggr).
\label{e109}
\end{gather*}

\section{Tau functions as matrix elements and bilinear identities}
Let $A$ be an operator  on $F$ such that
\begin{gather}
\left[A\otimes A, \Omega\right] = 0, \qquad \Omega=\sum
\psi_{\mathfrak k}^+\otimes\psi^-_{-\mathfrak k}. \label{commOm}
\end{gather}
Note that if $A\in GL_\infty$ then this $A$ satisf\/ies
(\ref{commOm}). So in the $n$-component case
\begin{gather*}
\Omega=\sum_{{\mathfrak k},i} \psi_{\mathfrak k}^{+(i)} \otimes
\psi^{-(i)}_{-{\mathfrak k}} = \sum_i\mathop{\rm Res}_z
\psi^{+(i)}(z) \otimes\psi^{-(i)}(z). \label{omega2}
\end{gather*}
Here $\mathop{\rm Res}_z \sum_i f_i z^i=f_{-1}$. Def\/ine the
following functions
\begin{gather}
\tau_{k_1,k_2,\ldots, k_n}^{m_1,m_2,\ldots,m_n}(t,s) = \langle
k_1,k_2,\ldots ,k_n|\widetilde{A}\,
|m_1,m_2,\ldots,m_n\rangle,\nonumber
\\
\widetilde{A}=\prod_{i=1}^n \Gamma_+^{(i)}(t^{(i)}) \, A \,
\prod_{j=1}^n\Gamma_-^{(j)}(-s^{(j)}). \label{tauA}
\end{gather}
Instead of $\tau_{k_1,k_2,\ldots, k_n}^{m_1,m_2,\ldots,m_n}(t,s)$
we shall write $\tau_\alpha^\beta(t,s)$ for
$\alpha=(\alpha_1,\alpha_2,\ldots, \alpha_n)$ and a similar
expression for $\beta$. We will also use
$|\alpha|=\alpha_1+\alpha_2+\cdots+ \alpha_n$,
 $|\alpha|_i=\alpha_1+\alpha_2+\cdots+\alpha_i$, $|\alpha|_0=0$ and
$\epsilon_j=(0,\ldots,0,1, 0,\ldots,0)$, a $1$ on the $j^{\rm th}$
place.

If the operator $A$ satisf\/ies \eqref{commOm}, then so does
$\widetilde{A}$. Following Okounkov \cite{O} we calculate in two
ways
\begin{gather}
\langle \alpha |\prod_{i=1}^n\Gamma_+^{(i)}({t}^{(i)}) \otimes
\langle \gamma| \prod_{k=1}^n\Gamma_+^{(k)}({s}^{(k)}) (A\otimes
A)\,\Omega \,\prod_{j=1}^n\Gamma_-^{(j)}(-{t'}^{(j)})
|\beta\rangle\otimes
\prod_{\ell=1}^n\Gamma_-^{(\ell)}(-{s'}^{(\ell)}) |\delta\rangle
\nonumber
\\ \qquad
=\!\langle \alpha
|\prod_{i=1}^n\!\Gamma_+^{(i)}({t}^{(i)})\!\otimes\! \langle
\gamma| \!\prod_{k=1}^n\!\Gamma_+^{(k)}({s}^{(k)}) \Omega
(A\!\otimes\! A)
\!\prod_{j=1}^n\!\Gamma_-^{(j)}(-{t'}^{(j)})|\beta\rangle\!\otimes\!
\prod_{\ell=1}^n\!\Gamma_-^{(\ell)}(-{s'}^{(\ell)})|\delta\rangle.\!\!\!
\label{bil}
\end{gather}
Clearly
\begin{gather*}
\psi^{+(m)}(z)\prod_{j=1}^n\Gamma_-^{(j)}(-{t'}^{(j)})
|\beta\rangle= Q_mz^{\alpha^{(m)}_{0}}\Gamma_{-}^{(m)}(
[z])\Gamma_{+}^{(m)}(-[z^{-1}])
\prod_{j=1}^n\Gamma_-^{(j)}(-{t'}^{(j)}) |\beta\rangle
\\ \qquad
{}=(-)^{|\beta|_{m-1}}z^{\beta_m}\gamma([z^{-1}],{t'}^{(m)})
\Gamma_{-}^{(m)}(
[z])\prod_{j=1}^n\Gamma_-^{(j)}(-{t'}^{(j)})|\beta+\epsilon_m\rangle
\\ \qquad
{}=(-)^{|\beta|_{m-1}}z^{\beta_m}\gamma([z^{-1}],{t'}^{(m)})
\Gamma_-^{(m)}(-{t'}^{(m)}+[z])\prod_{j\ne
m}^n\Gamma_-^{(j)}(-{t'}^{(j)}) |\beta+\epsilon_m\rangle
\end{gather*}
and in a similar way
\begin{gather*}
\psi^{-(m)}(z)\prod_{\ell=1}^n\Gamma_-^{(\ell)}(-{s'}^{(1)})|\delta\rangle
\\ \qquad
{}=(-)^{|\delta|_{m-1}}z^{-\delta_m}\gamma([z^{-1}],-{s'}^{(m)})
\Gamma_-^{(m)}(-{s'}^{(m)}-[z]) \prod_{\ell\ne
m}\Gamma_-^{(\ell)}(-{s'}^{(\ell)}) |\delta-\epsilon_m\rangle.
\end{gather*}
Also
\begin{gather*}
\langle \alpha|\prod_{i=1}^n\Gamma_+^{(i)}({t}^{(i)})
\psi^{+(m)}(z)= \langle
\alpha|\prod_{i=1}^n\Gamma_+^{(i)}({t}^{(i)})
Q_mz^{\alpha^{(m)}_{0}}\Gamma_{-}^{(m)}( [z])\Gamma_{+}^{(m)}(-
[z^{-1}])
\\ \qquad
{}=(-)^{|\alpha|_{m-1}}z^{\alpha_m-1}\gamma([z],{t}^{(m)})\langle
\alpha-\epsilon_m| \prod_{i=1}^n\Gamma_+^{(i)}({t}^{(i)})
\Gamma_{+}^{(m)}(- [z^{-1}])
\\ \qquad
{}=(-)^{|\alpha|_{m-1}}z^{\alpha_m-1}\gamma([z],{t}^{(m)})\langle
\alpha-\epsilon_m| \Gamma_+^{(m)}({t}^{(m)}-[z^{-1}])\prod_{i\ne
m}\Gamma_+^{(i)}({t}^{(i)})
\end{gather*}
and
\begin{gather*}
\begin{split}
\langle \gamma| \prod_{k=1}^n&\Gamma_+^{(k)}({s}^{(k)})
\psi^{-(m)}(z)\\
&=(-)^{|\gamma|_{m-1}}z^{-\gamma_m-1}\gamma({s}^{(m)},-[z])\langle
\gamma+\epsilon_m| \Gamma_+^{(m)}({s}^{(m)}+[z^{-1}])\prod_{k\ne
m}\Gamma_+^{(k)}({s}^{(k)}).
\end{split}
\end{gather*}
Using this we rewrite (\ref{bil}):
\begin{gather}
\mathop{\rm Res}_z
\sum_{m=1}^n(-)^{|\beta+\delta|_{m-1}}z^{\beta_m-\delta_m}
\gamma([z^{-1}],{t'}^{(m)}-{s'}^{(m)})\tau_{\alpha}^{\beta+
\epsilon_m}(t^{(i)},{t'}^{(j)}-\delta_{jm}[z])\nonumber
\\ \qquad
{}\times\tau_{\gamma}^{\delta-\epsilon_m}(s^{(k)},{s'}^{(\ell)}+\delta_{\ell
m}[z])= \mathop{\rm Res}_z \sum_{m=1}^n
(-)^{|\alpha+\gamma|_{m-1}}z^{\alpha_m-\gamma_m-2}
\gamma([z],{t}^{(m)}-{s}^{(m)})\nonumber
\\ \qquad
{}\times\tau_{\alpha-\epsilon_m}^{\beta}(t^{(i)}-\delta_{im}[z^{-1}],{t'}^{(j)})
\tau_{\gamma+\epsilon_m}^{\delta}(s^{(k)}+\delta_{km}[z^{-1}],{s'}^{(\ell)}).
\label{bil2}
\end{gather}
For $n=1$ this is the Toda lattice hierarchy of Ueno and Takasaki
\cite{UT}.

We now rewrite the left-hand side of (\ref{bil2}) to obtain a more
familiar form. For this we replace~$z$ by $z^{-1}$ and write $u$,
resp. $v$ for $t'$, resp. $s'$, we thus obtain the following
bilinear identity.
\begin{proposition}
\label{TODA} The tau functions satisfy the following identity:
\begin{gather}
\mathop{\rm Res}_z
\sum_{m=1}^n(-)^{|\beta+\delta|_{m-1}}z^{\delta_m-\beta_m-2}
\gamma([z],{u}^{(m)}-{v}^{(m)})
\tau_{\alpha}^{\beta+\epsilon_m}(t^{(i)},{u}^{(j)}
-\delta_{jm}[z^{-1}])\nonumber
\\ \qquad
{}\times\tau_{\gamma}^{\delta-\epsilon_m}(s^{(k)},{v}^{(\ell)}+\delta_{\ell
m}[z^{-1}])= \mathop{\rm Res}_z \sum_{m=1}^n
(-)^{|\alpha+\gamma|_{m-1}}z^{\alpha_m-\gamma_m-2}
\gamma([z],{t}^{(m)}-{s}^{(m)})\nonumber
\\ \qquad
{}\times\tau_{\alpha-\epsilon_m}^{\beta}(t^{(i)}-\delta_{im}[z^{-1}],{u}^{(j)})
\tau_{\gamma+\epsilon_m}^{\delta}(s^{(k)}+\delta_{km}[z^{-1}],{v}^{(\ell)}).
\label{bil3}
\end{gather}
\end{proposition}

From the commutation relations (\ref{comm}) one easily deduces the following
\begin{proposition}
\label{prop2.1} Let $\lambda=+,-$,  and $a_j\in\mathbb{C}$, for
$1\le j\le n$.

{\rm (a)} The operator $A=\sum\limits_{j=1}^n
a_j\psi^{\lambda(j)}(z)$ satisfies \eqref{commOm}.

{\rm (b)} Let $f_j(z)=\sum\limits_i f_j^i z^i$, for $1\le j\le n$,
then the operator $A=\sum\limits_{j=1}^n \mathop{\rm Res}_z \,
f_j(z)\psi^{\lambda(j)}(z)$ satis\-fies~\eqref{commOm}.
\end{proposition}

Another way of obtaining solutions  is as follows. We sketch the
case $n=2$ (see e.g.~\cite{HO}) which is related to a two matrix
model. Let $d\mu (x,y)$ be a measure (in general complex),
supported either on a f\/inite set of products of curves in the
complex $x$ and $y$ planes or, alternatively, on a~domain in the
complex $z$ plane, with identif\/ications $x=z$, $y=\bar z$. Then,
for  each $1\le j,k\le n$ and $\lambda,\nu=+,-$ the operator
$A=e^B$ with
\begin{gather*}
B=\int\psi^{\lambda(j)}(x)\psi^{\nu(k)}(y)\,d\mu (x,y)
\end{gather*}
satisfy (\ref{commOm}). If one chooses $j=1,\ k=2$ and
$\lambda=+,\ \nu=-$ and def\/ines for the above $A$
\begin{gather*}
\tau_{\alpha_1,\alpha_2}^{\alpha_1+N, \alpha_2-N}(t,u)= \langle
\alpha_1,\alpha_2 |\Gamma_+^{(1)}(t^{(1)}) \Gamma_+^{(2)}(t^{(2)})
e^B \Gamma_-^{(1)}(-u^{(1)})\Gamma_-^{(2)}(-u^{(2)})|\alpha_1+N,
\alpha_2-N\rangle
\\ \qquad
{}=\frac{1}{N!}\langle \alpha_1,\alpha_2 |\Gamma_+^{(1)}(t^{(1)})
\Gamma_+^{(2)}(t^{(2)})  B^N \,
\Gamma_-^{(1)}(-u^{(1)})\Gamma_-^{(2)}(-u^{(2)})|\alpha_1+N,
\alpha_2-N\rangle,
\end{gather*}
then these tau-functions satisfy~\eqref{bil3}.

\section[Wave functions and pseudo-differential equations]{Wave functions and pseudo-dif\/ferential equations}

We will now rewrite the equations (\ref{bil3}) in another form.
Note that for
\begin{gather}
{}|\beta-\alpha|=|\delta-\gamma|=j\qquad\mbox{for f\/ixed}\quad
j\in\mathbb Z \label{j}
\end{gather}
these are the equations of the $2n$-component KP hierarchy, see
\cite{KL}. One obtains equation (66) of \cite{KL} if one  chooses
$t^{(m+n)}=u^{(m)}$,  $s^{(m+n)}=v^{(m)}$  and $\tau_\alpha^\beta
(t)=(i)^{|\alpha|^2}\tau_{(\alpha,-\beta)}(t)$, where
\begin{gather*}
\tau_{(\alpha,-\beta)}(t)=
\tau_{(\alpha_1,\alpha_2,\ldots,\alpha_n,-\beta_1,-\beta_2,\ldots,
-\beta_n)}(t)
\end{gather*}
are the $2n$-component KP tau-functions, viz for
$(\alpha,-\beta)=\rho$ and $(\gamma,-\delta)=\sigma$ they satisfy
the $2n$-component KP equations:
\begin{gather*}
\mathop{\rm Res}_z
\sum_{m=1}^{2n}(-)^{|\rho+\sigma|_{m-1}}z^{\rho_m-\sigma_m-2}
\gamma([z],{t}^{(m)}-{s}^{(m)})\tau_{\rho-\epsilon_m}(t^{(i)}-\delta_{im}[z^{-1}])
\\ \qquad
{}\times \tau_{\sigma+\epsilon_m}(s^{(k)}+\delta_{km}[z^{-1}])=0.
\end{gather*}
In \cite{KL} one showed that one can rewrite these equations to
get $2n\times 2n$ matrix wave functions. Here we want to obtain
two $n\times n$ matrix wave functions. {\it We assume from now on
that \eqref{j} holds for $j=0$.} Denote ($s=0,1$):
\begin{gather}
P^{\pm(s)}(\alpha,\beta,t,u,\pm z)= (
P^{\pm(s)}(\alpha,\beta,t,u,\pm z)_{k\ell})_{1\le k,\ell\le
n},\nonumber
\\
P^{\pm(0)}(\alpha,\beta,t,u,\pm z)_{k\ell}=(-)^{|\epsilon_k|_{\ell
-1}}z^{\delta_{k\ell}-1}
\frac{\tau_{\alpha\pm(\epsilon_k-\epsilon_\ell)}^\beta(t^{(i)}\mp
\delta_{i\ell}[z^{-1}],u^{(j)})}{\tau_{\alpha}^\beta(t^{(i)},u^{(j)})},\nonumber
\\
P^{\pm(1)}(\alpha,\beta,t,u,\pm
z)_{k\ell}=z^{-1}\frac{\tau_{\alpha\pm\epsilon_k}^{\beta\pm
\epsilon_\ell} (t^{(i)},u^{(j)}\mp
\delta_{j\ell}[z^{-1}])}{\tau_{\alpha}^\beta(t^{(i)},u^{(j)})},\nonumber
\\
R^{\pm(s)}(\alpha,\beta,\pm z)={\rm
diag}(R^{\pm(s)}(\alpha,\beta,\pm z)_\ell),\nonumber
\\
R^{\pm(0)}(\alpha,\beta,\pm z)_\ell=(-)^{|\alpha|_{\ell -1}}z^{\pm
\alpha_\ell},\nonumber
\\
R^{\pm(1)}(\alpha,\beta,\pm z)_\ell=(-)^{|\beta|_{\ell
-1}}z^{\mp\beta_\ell},\nonumber
\\
Q^\pm(t,\pm z)={\rm diag}(\gamma([z],\pm t^{(1)}),\gamma([z],\pm
t^{(2)}), \ldots,\gamma([z],\pm t^{(n)})). \label{wave}
\end{gather}
Replace $t^{(a)}_1$, $s^{(a)}_1$, $u^{(a)}_1$, $v^{(a)}_1$ by
$t^{(a)}_1+x_0$, $s^{(a)}_1+y_0$, $u^{(a)}_1+x_1$, $v^{(a)}_1+y_1$
then some of the above functions also depend on $x_0$ and $x_1$ we
will add these variables to these functions and write e.g.
$P^{\pm(0)}(\alpha,\beta,x,t,u,\pm z)$ for
$P^{\pm(0)}(\alpha,\beta,t_i^{(j)}+\delta_{i1}x_0,u,\pm z)$.
Introduce for $k=0,1$ the following dif\/ferential symbols
$\partial_k={\partial}/{\partial x_k}$ ${\partial
'}_k={\partial}/{\partial y_k}$. Introduce the wave functions,
here $x$ is short hand notation for $x=(x_0,x_1)$
\begin{gather}
\Psi^{\pm(0)}(\alpha,\beta,x,t,u,z)=P^{\pm(0)}(\alpha,\beta,x,t,u,\pm
z) R^{\pm(0)}(\alpha,\beta,\pm z)Q^\pm(t,\pm z)e^{\pm
x_0z}\nonumber
\\ \phantom{\Psi^{\pm(0)}(\alpha,\beta,x,t,u,z)}{}
{}=P^{\pm(0)}(\alpha,\beta,x,t,u,\partial_0)R^{\pm(0)}(\alpha,\beta,\partial_0)
Q^\pm(t,\partial_0)e^{\pm x_0z},\nonumber
\\
\Psi^{\pm(1)}(\alpha,\beta,x,t,u,z)=P^{\pm(1)}(\alpha,\beta,x,t,u,\pm
z) R^{\pm(1)}(\alpha,\beta,\pm z)Q^\pm(u,\pm z)e^{\pm
x_1z}\nonumber
\\ \phantom{\Psi^{\pm(1)}(\alpha,\beta,x,t,u,z)}{}
{}=
P^{\pm(1)}(\alpha,\beta,x,t,u,\partial_1)R^{\pm(1)}(\alpha,\beta,\partial_1)
Q^\pm(u,\partial_1)e^{\pm x_1z}. \label{wwave}
\end{gather}
Then (\ref{bil3}) leads to
\begin{gather}
\mathop{\rm Res}_z
\Psi^{+(0)}(\alpha,\beta,x,t,u,z)\Psi^{-(0)}(\gamma,\delta,y,s,v,z)^T\nonumber
\\ \qquad
{}=\mathop{\rm
Res}_z\Psi^{+(1)}(\alpha,\beta,x,t,u,z)\Psi^{-(1)}(\gamma,\delta,y,s,v,z)^T.
\label{bil4}
\end{gather}
One can also deduce the following 6 equations, for a proof see the
appendix A:
\begin{gather}
\label{third} P^{+(0)}(\alpha,\beta,x_0,x_1,t,u,\partial_0)^{-1}=
P^{-(0)}(\alpha,\beta,x_0,y_1,s,v,\partial_0)^*,
\\
\label{invP1}
P^{+(1)}(\alpha,\beta,x_0,x_1,t,u,\partial_1)^{-1}\!=\!
S(\partial_1) P^{-(1)}\Biggl(\!\alpha\!+\!\sum_{i=1}^n
\epsilon_i,\beta\!+\!\sum_{i=1}^n
\epsilon_i,x_0,x_1,t,u,\partial_1\biggr)^* S(\partial_1),\!\!\!
\\
\frac{\partial
P^{+(0)}(\alpha,\beta,x_0,x_1,t,u,\partial_0)}{\partial t_j^{(a)}}
\nonumber
\\ \qquad
{}=-(P^{+(0)}(\alpha,\beta,x_0,x_1,t,u,\partial_0)E_{aa}\partial_0^j
P^{+(0)}(\alpha,\beta,x_0,x_1,t,u,\partial_0)^{-1})_- \nonumber
\\ \qquad
{}\times P^{+(0)}(\alpha,\beta,x_0,x_1,t,u,\partial_0),
\label{Sato}
\\
\frac{\partial P^{+(1)}(\alpha,\beta,x,t,u,\partial_1)}{\partial
u_j^{(a)}} =-(P^{+(1)}(\alpha,\beta,x,t,u,\partial_1)
\partial_1^j E_{aa}P^{+(1)}(\alpha,\beta,x,t,u,\partial_1)^{-1})_{<1}\nonumber
\\  \qquad
{}\times P^{+(1)}(\alpha,\beta,x,t,u,\partial_1). \label{Sato4}
\end{gather}
And
\begin{gather}
\sum_{j=1}^\infty\frac{\partial
P^{+(0)}(\alpha,\beta,x,t,u,\partial_0)}{\partial
u_j^{(a)}}z^{-j-1}\nonumber
\\ \qquad
{}=P^{+(1)}(\alpha,\beta,x,t,u,z)E_{aa}\partial_0^{-1}
P^{-(1)}(\alpha,\beta,x,t,u,-z)^TP^{+(0)}(\alpha,\beta,x,t,u,\partial_0),
\label{sato22}
\\
\Biggl(E_{aa} +\sum_{j=1}^\infty\frac{\partial }{\partial
t_j^{(a)}}z^{-j}\Biggr)(P^{+(1)}(\alpha,\beta,x,t,u,\partial_1))=
P^{+(0)}(\alpha,\beta,x,t,u,z)E_{aa}S(\partial_1)^{-1}\nonumber
\\ \qquad
{}\times P^{-(0)}\Biggl(\alpha+\sum_{i=1}^n
\epsilon_i,\beta+\sum_{i=1}^n
\epsilon_i,x,s,v,-z\Biggr)^TS(\partial_1)
P^{+(1)}(\alpha,\beta,x,t,u,\partial_1), \label{Sato33}
\end{gather}
where
\begin{gather*}
S(\partial)=\sum_{i=1}^n (-)^{i+1}E_{ii}\partial. \label{S}
\end{gather*}

Recall that $(P(x)\partial^k)^*= (-\partial)^k\cdot P(x)^T$.

\section{First reduction and the generalized AKNS model}

Assume  from now on that our tau functions satisfy
\begin{gather*}
\tau^{\beta-\sum\limits_j\epsilon_j}_{\alpha-\sum\limits_j\epsilon_j}=(-)^{\sum\limits_{i=1}^{n-1}
|\beta-\alpha|_i}c\tau_{\alpha}^{\beta},\qquad\mbox{where}\quad
0\ne c\in\mathbb{C}.
\end{gather*}
This holds e.g. when
\begin{gather}
Q_1Q_2\cdots Q_nA=cAQ_1Q_2\cdots Q_n. \label{A}
\end{gather}
%  but also for the fermionic operators that appear in (5.24) of \cite{LM}.
Substituting this in (\ref{wave}), gives
\begin{gather}
P^{\pm
(i)}\Biggl(\alpha-\sum_j\epsilon_j,\beta-\sum_j\epsilon_j,t,u,\partial_i\Biggr)
=\sum_{k=1}^n (-)^kE_{kk}P^{\pm
(i)}(\alpha,\beta,t,u,\partial_i)\sum_{k=1}^n (-)^kE_{kk}.
\label{invP11}
\end{gather}
Now from the second equation of (\ref{second}) and (\ref{Sato}) we
deduce that
\begin{gather}
\sum_{j=1}^n\frac{\partial P^{+(0)}(\alpha, \beta,x_0,
x_1,t,u,\partial_0)}{\partial t_1^{(j)}}=0. \label{1}
\end{gather}
Hence this implies that
\begin{gather}
[\partial_0,P^{+(0)}(\alpha, \beta,x_0,x_1,t,u,\partial_0)]=0.
\label{2}
\end{gather}
In a similar way we deduce from (\ref{fifth}) and (\ref{Sato4})
that
\begin{gather}
\sum_{j=1}^n\frac{\partial P^{+(1)}(\alpha,
\beta,x_0,x_1,t,u,\partial_1)}{\partial u_1^{(j)}}=0. \label{3}
\end{gather}
Hence also
\begin{gather}
[\partial_1,P^{+(1)}(\alpha, \beta,x_0,x_1,t,u,\partial_1)]=0.
\label{4}
\end{gather}

\noindent {\bf Remark.} Note that the above does not imply that
both
\begin{gather*}
[\partial_1,P^{+(0)}(\alpha, \beta,x_0,x_1,t,u,\partial_0)]=0
\qquad\mbox{and}\qquad [\partial_0,P^{+(1)}(\alpha,
\beta,x_0,x_1,t,u,\partial_1)]=0.
\end{gather*}
Next, using this reduction and (\ref{P+-}) then (\ref{SSato3})
turns into
\begin{gather}
\frac{\partial P^{+(1)}(\alpha, \beta,x,t,u,\partial_1)}{\partial
x_0}\nonumber
\\ \qquad
{}=\left(P^{+(0)}_1(\alpha, \beta,x,t,u)\partial_1^{-1}-
\partial_1^{-1}P^{+(0)}_1(\alpha, \beta,x,t,u)\right)\partial_1
P^{+(1)}(\alpha, \beta,x,t,u,\partial_1)\nonumber
\\ \qquad
{}=\sum_{j=1}^\infty (-)^{j+1}\frac{\partial^j P^{+(0)}_1 (\alpha,
\beta,x,t,u)}{\partial x_1^j}P^{+(1)}(\alpha,
\beta,x,t,u,\partial_1)\partial_1^{-j}. \label{SSSato3}
\end{gather}
Now, use (\ref{invP1}) to see that (\ref{***}) turns into
\begin{gather}
\frac{\partial P^{+(0)}(\alpha, \beta,x,t,u,\partial_0)}{\partial
x_1}\nonumber
\\ \qquad{}=
P^{+(1)}_1(\alpha, \beta,x,t,u)\partial_0^{-1} P^{(1)}_1(\alpha,
\beta,x,t,u,)^{-1} P^{+(0)}(k_1,k_2,x,t,u,\partial_0).
\label{****}
\end{gather}
Taking the coef\/f\/icient of $\partial_0^{-1}$ of this equation
we thus get
\begin{gather*}
\frac{\partial P^{+(0)}_1(\alpha, \beta,x,t,u)}{\partial x_1}=I
\end{gather*}
and hence (\ref{SSSato3}) turns into
\begin{gather*}
\frac{\partial P^{+(1)}(\alpha, \beta,x,t,u,\partial_1)}{\partial
x_0}=P^{+(1)}(\alpha, \beta,x,t,u,\partial_1)\partial_1^{-1}.
\label{sato}
\end{gather*}
In particular
\begin{gather*}
\frac{\partial P^{+(1)}_1(\alpha, \beta,x,t,u)}{\partial x_0}=0.
\end{gather*}
Substituting this in (\ref{****}) we obtain
\begin{gather}
\frac{\partial P^{+(0)}(\alpha, \beta,x,t,u,\partial_0)}{\partial
x_1}=P^{+(0)}(\alpha, \beta,x,t,u,\partial_0)\partial_0^{-1}.
\label{ssato}
\end{gather}
Note f\/irst that from (\ref{ssato})  one deduces that
\begin{gather*}
\frac{\partial P^{+(0)}(\alpha,
\beta,x,t,u,\partial_0)^{-1}}{\partial x_1}
\\ \qquad
{}=-P^{+(0)}(\alpha, \beta,x,t,u,\partial_0)^{-1} \frac{\partial
P^{+(0)}(\alpha, \beta,x,t,u,\partial_0)}{\partial x_1}
P^{+(0)}(\alpha, \beta,x,t,u,\partial_0)^{-1}\partial_0^{-1}
\\ \qquad
{}=-P^{+(0)}(\alpha, \beta,x,t,u,\partial_0)^{-1}P^{+(0)}(\alpha,
\beta,x,t,u,\partial_0)
\partial_0^{-1}P^{+(0)}(\alpha, \beta,x,t,u,\partial_0)^{-1}
\\ \qquad
{}=- P^{+(0)}(\alpha, \beta,x,t,u,\partial_0)^{-1}\partial_0^{-1}.
\end{gather*}
 Clearly also
 \begin{gather*}
 \frac{\partial  P^{+(1)}(\alpha, \beta,x,t,u,\partial_1)^{-1}}{\partial
x_0}=-P^{+(1)}(\alpha, \beta,x,t,u,\partial_1)^{-1}
\partial_1^{-1}.
\end{gather*}
Using (\ref{third}) we rewrite (\ref{Sato3}):
\begin{gather}
\frac{\partial P^{+(1)}(\alpha, \beta,x,t,u,\partial_1)}{\partial
t_j^{(a)}}= \mathop{\rm Res}_z P^{+(0)}(\alpha,
\beta,x,t,u,z)z^{j-1}E_{aa}\nonumber
\\ \qquad
{}\times\partial_1^{-1}P^{+(0)}(\alpha,
\beta,x,t,u,z)^{-1}\partial_1 P^{+(1)}(\alpha,
\beta,x,t,u,\partial_1)\nonumber
\\ \qquad
{}=\mathop{\rm Res}_z P^{+(0)}(\alpha,
\beta,x,t,u,z)z^{j-1}E_{aa}\nonumber
\\ \qquad
{}\times\sum_{i=0}^\infty (-)^i \frac{\partial^i P^{+(0)}(\alpha,
\beta,x,t,u,z)^{-1}}{\partial x_1^i}
\partial_1^{-i}P^{+(1)}(\alpha, \beta,x,t,u,\partial_1)\nonumber
\\ \qquad
{}=\mathop{\rm Res}_z P^{+(0)}(\alpha,
\beta,x,t,u,z)E_{aa}\nonumber
\\ \qquad
{}\times P^{+(0)}(\alpha, \beta,x,t,u,z)^{-1} \sum_{i=0}^\infty
z^{j-i-1}P^{+(1)}(\alpha, \beta,x,t,u,\partial_1)
\partial_1^{-i}\nonumber
\\ \qquad
{}=\mathop{\rm Res}_z \sum_{i=0}^j  z^{j-i-1}P^{+(0)}(\alpha,
\beta,x,t,u,z)E_{aa}\nonumber
\\ \qquad
{}\times P^{+(0)}(\alpha, \beta,x,t,u,z)^{-1}P^{+(1)}(\alpha,
\beta,x,t,u,\partial_1)
\partial_1^{-i}.
\label{sato3}
\end{gather}
N.B. This summation starts with 0.

In particular
\begin{gather}
\frac{\partial P^{+(1)}(\alpha, \beta,x,t,u,\partial_1)}{\partial
t_j^{(1)}} +\frac{\partial P^{+(1)}(\alpha,
\beta,x,t,u,\partial_1)}{\partial t_j^{(2)}} =P^{+(1)}(\alpha,
\beta,x,t,u,\partial_1)\partial_1^{-j}. \label{sato4}
\end{gather}
In a similar way we deduce, using (\ref{invP1}), from
(\ref{sato2}):
\begin{gather}
\frac{\partial P^{+(0)}(\alpha, \beta,x,t,u,\partial_0)}{\partial
u_j^{(a)}}= \mathop{\rm Res}_z \sum_{i=1}^j  z^{j-i-1}
P^{+(1)}(\alpha, \beta,x,t,u,z)E_{aa}\nonumber
\\ \qquad
{}\times P^{+(1)}(\alpha, \beta,x,t,u,z)^{-1} P^{+(0)}(\alpha,
\beta,x,t,u,\partial_0)\partial_0^{-i}. \label{sato5}
\end{gather}
N.B. This summation starts with 1.

In particular
\begin{gather}
\sum_{k=1}^n\frac{\partial P^{+(0)}(\alpha,
\beta_2,x,t,u,\partial_0)} {\partial u_j^{(k)}}=P^{+(0)}(\alpha,
\beta,x,t,u,\partial_0)\partial_0^{-j}.
 \label{sato6}
\end{gather}

We will now combine (\ref{Sato}), (\ref{Sato4}),
(\ref{1})--(\ref{4}) and (\ref{sato3})--(\ref{sato6}). For this
purpose we replace~$\partial_0$ by the loop variable $z$ and
$\partial_1$ by the loop variable $z^{-1}$. We write
\begin{gather}
P^{(0)}(\alpha, \beta,x,t,u,z)= P^{+(0)}(\alpha,
\beta,x_0,x_1,t,u,z)\nonumber \qquad\mbox{and}
\\
P^{(1)}(\alpha, \beta,x,t,u,z)= z^{-1}
P^{+(1)}(\alpha,\beta,x_0,x_1,t,u,z^{-1}) \label{loop0}
\end{gather}
and  thus obtain for f\/ixed $\alpha$ and  $\beta$
($P^{(j)}(x,t,u,z)=P^{(j)}(\alpha, \beta,x,t,u,z)$):
\begin{gather}
\frac{\partial P^{(0)}(x,t,u,z)}{\partial x_0}=0,\nonumber \qquad
\frac{\partial P^{(0)}(x,t,u,z)}{\partial
x_1}=P^{(0)}(x,t,u,z)z^{-1},\nonumber
\\
\frac{\partial P^{(1)}(x,t,u,z)}{\partial
x_0}=P^{(1)}(x,t,u,z)z,\nonumber \qquad \frac{\partial
P^{(1)}(x,t,u,z)}{\partial x_1}=0,\nonumber
\\
\frac{\partial P^{(0)}(x,t,u,z)}{\partial t_j^{(a)}}=
-(P^{(0)}(x,t,u,z)E_{aa}P^{(0)}(x,t,u,z)^{-1}z^j)_-P^{(0)}(x,t,u,z),\nonumber
\\
\frac{\partial P^{(0)}(x,t,u,z)}{\partial u_j^{(a)}}
=(P^{(1)}(x,t,u,z)E_{aa}P^{(1)}(x,t,u,z)^{-1}z^{-j})_-P^{(0)}(x,t,u,z),\nonumber
\\
\frac{\partial P^{(1)}(x,t,u,z)}{\partial
t_j^{(a)}}=(P^{(0)}(x,t,u,z)E_{aa} P^{(0)}(x,t,u,z)^{-1}
z^j)_+P^{(1)}(x,t,u,z),\nonumber
\\
\frac{\partial P^{(1)}(x,t,u,z)}{\partial
u_j^{(a)}}=-(P^{(1)}(x,t,u,z)
E_{aa}P^{(1)}(x,t,u,z)^{-1}z^{-j})_{+}P^{(1)}(x,t,u,z).
\label{loop}
\end{gather}
Which are the generalized AKNS equations (2.5)--(2.10) of
\cite{AGZ}. Write
\begin{gather*}
P^{(0)}(x,t,u,z)=I+\sum_{i=1}^\infty P_i(x,t,u)z^{-i},\qquad
P^{(1)}(x,t,u,z)=\sum_{i=0}^\infty M_i(x,t,u)z^i
\end{gather*}
and from now on in this section
\begin{gather*}
\partial_j=\frac{\partial}{\partial t_1^{(j)}},
\qquad \partial_{-j}=\frac{\partial}{\partial u_1^{(j)}}.
\end{gather*}
We thus obtain the following equations:
\begin{gather}
\partial_j P_1=[P_1,E_{jj}]P_1+ [E_{jj},P_2],\nonumber
\qquad
\partial_{-j} P_1=M_0E_{jj} M_0^{-1},\nonumber
\\[1ex]
\partial_j M_0=[P_1, E_{jj}]M_0,\nonumber
\qquad
\partial_j M_1=E_{jj}M_0+[P_1, E_{jj}]M_1,\nonumber
\\[1ex]
\partial_{-j} M_0=M_0[E_{jj},M_0^{-1}M_1 ],
\qquad
\partial_{-j} M_1=M_0[E_{jj},M_0^{-1}M_2 ].
\label{PM}
\end{gather}
From this one easily deduces the following equations for $M_0$:
\begin{gather*}
\partial_i(M_0^{-1}\partial_{-j}M_0)=[E_{jj},M_0^{-1}E_{ii}M_0],\nonumber
\\[1ex]
\partial_{-i}(M_0E_{jj}M_0^{-1})=\partial_{-j}(M_0E_{ii}M_0^{-1}),\nonumber
\\[1ex]
\partial_{i}(M_0^{-1}E_{jj}M_0)=\partial_{j}(M_0^{-1}E_{ii}M_0).
\label{CV1}
\end{gather*}
Note that if we def\/ine $\bar P_i=M_0^{-1}M_i$, then from
(\ref{PM}) we get:
\begin{gather}
\partial_j P_1=[P_1,E_{jj}]P_1+ [E_{jj},P_2],
\qquad
\partial_{-j} P_1=M_0E_{jj} M_0^{-1},
\qquad
\partial_j \bar P_1=M_0^{-1}E_{jj} M_0,\nonumber
\\[1ex]
\partial_{j} \bar P_1=[\bar P_1,E_{jj}]\bar P_1+ [E_{jj},\bar P_2],
\qquad
\partial_j M_0=[P_1, E_{jj}]M_0,
\qquad
\partial_{-j} M_0=M_0[E_{jj},\bar P_1 ].
\label{PM2}
\end{gather}

Note that for $x_0=x_1=0$ we have
\begin{gather}
(M_0(\alpha,\beta,t,u))_{kl}=
\frac{\tau^{\beta+\epsilon_\ell}_{\alpha+\epsilon_k}(t,u)}
{\tau^{\beta}_{\alpha}(t,u)}, \qquad
(M_1(\alpha,\beta,t,u))_{kl}=-\frac{\partial_{-\ell}
(\tau^{\beta+\epsilon_\ell}_{\alpha+\epsilon_k}(t,u))}
{\tau^{\beta}_{\alpha}(t,u)},\nonumber
\\
(M_2(\alpha,\beta,t,u))_{kl}
=\frac{1}{2}\frac{(\partial_{-\ell}^2-\partial_{u_2^{(\ell)}})
(\tau^{\beta+\epsilon_\ell}_{\alpha+\epsilon_k}(t,u))}
{\tau^{\beta}_{\alpha}(t,u)},\nonumber
\\
(P_1(\alpha,\beta,t,u))_{kl}=
\begin{cases}\displaystyle
(-)^{|\epsilon_k|_{\ell-1}}
\frac{\tau^{\beta}_{\alpha+\epsilon_k-\epsilon_\ell}(t,u)}
{\tau^{\beta}_{\alpha}(t,u)}\qquad\mbox{if}\qquad k\ne
\ell,\nonumber
\\
\displaystyle -\frac{\partial_{\ell}(\tau^{\beta}_{\alpha}(t,u))}
{\tau^{\beta}_{\alpha}(t,u)}\qquad\mbox{if}\qquad k=\ell,
\end{cases}
\\
(P_2(\alpha,\beta,t,u))_{kl}=
\begin{cases}\displaystyle
-(-)^{|\epsilon_k|_{\ell-1}}
\frac{\partial_\ell\left(\tau^{\beta}_{\alpha+\epsilon_k-\epsilon_\ell}(t,u)
\right)} {\tau^{\beta}_{\alpha}(t,u)}\qquad\mbox{if}\qquad k\ne
\ell,
\\
\displaystyle
\frac{1}{2}\frac{\left(\partial_{\ell}^2-\partial_{t_2^{\ell}}\right)
\left(\tau^{\beta}_{\alpha}(t,u)\right)}
{\tau^{\beta}_{\alpha}(t,u)}\qquad\mbox{if}\qquad k=\ell.
\end{cases}
\label{ttau}
\end{gather}

\section[AKNS and the two-component Camassa-Holm model]{AKNS and the two-component Camassa--Holm model}
We still assume that $A$ satisf\/ies (\ref{A}), hence that we have
the reduction of the previous section. We consider the case $n=2$
and def\/ine
\begin{gather}
y=\frac{t^{(1)}_1-t^{(2)}_1}2,\qquad \bar
y=\frac{t^{(1)}_1+t^{(2)}_1}2, \qquad s=
2u^{(1)}_1-2u^{(2)}_1,\qquad \bar s= 2u^{(1)}_1+2u^{(2)}_1,
\label{ys}
\end{gather}
then
\begin{gather*}
\frac{\partial}{\partial_y}=\partial_1-\partial_2,\qquad
\frac{\partial}{\partial_s}=\frac{1}{4}
\left(\partial_{-1}-\partial_{-2}\right).
\end{gather*}
Now Let $E=E_{11}-E_{22}$ and def\/ine
\begin{gather*}
\psi(z)=P^{(0)}(0,t,u,z)\begin{pmatrix}
e^{yz}&0\\
0&e^{-yz}
\end{pmatrix}
\end{gather*}
then (\ref{loop}) turns into:
\begin{gather*}
\frac{\partial\psi(z)}{\partial y}= (zE+[P_1,E])\psi(z)
=\left(\begin{pmatrix}
z&0\\
0&-z
\end{pmatrix}+\begin{pmatrix}
0&q\\
r&0
\end{pmatrix}\right)\psi(z),
\\
\frac{\partial\psi(z)}{\partial
s}=\frac{z^{-1}}{4}M_0EM_0^{-1}\psi(z) =z^{-1}\begin{pmatrix}
A&B\\
C&-A
\end{pmatrix}\psi(z),
\end{gather*}
where
\begin{gather*}
q=-2(P_1)_{12},\qquad r=2(P_1)_{21}, \qquad A=\frac{1}{4\det
M_0}((M_0)_{11}(M_0)_{22}+(M_0)_{12}(M_0)_{21}),
\\
B=\frac{-1}{2\det M_0}(M_0)_{11}(M_0)_{12}, \qquad
C=\frac{1}{2\det M_0}(M_0)_{22}(M_0)_{21}.
\end{gather*}
Now (\ref{PM}) turns into:
\begin{gather}
\frac{\partial P_1}{\partial y}\!=\![P_1,E]P_1\!+\! [E,P_2]
=\begin{pmatrix}
0&q\\
r&0
\end{pmatrix}P_1\!+\![E,P_2],
\quad \frac{\partial P_1}{\partial s}\!=\!\frac{1}{4}M_0E M_0^{-1}
=\begin{pmatrix}
A&B\\
C&-A
\end{pmatrix},\nonumber
\\
\frac{\partial M_0}{\partial y}=[P_1, E]M_0 =\begin{pmatrix}
0&q\\
r&0
\end{pmatrix}M_0,
\quad \frac{\partial M_1}{\partial y}=EM_0+[P_1, E]M_1=EM_0+
\begin{pmatrix}
0&q\\
r&0
\end{pmatrix}M_1,\nonumber
\\
\frac{\partial M_0}{\partial s}=\frac{1}{4}M_0[E,M_0^{-1}M_1 ],
\qquad \frac{\partial M_1}{\partial
s}=\frac{1}{4}M_0[E,M_0^{-1}M_2 ]. \label{PM1}
\end{gather}
%One thus has
%\begin{gather}
%\label{qrs}
%\frac{\partial q}{\partial s}=-2B,\qquad \qquad\frac{\partial r}{\partial
%%s}=2C
%\end{gather}
%and
%\begin{gather}
%\label{ABCy}
%\frac{\partial A}{\partial y}=qC-rB,\qquad
%\frac{\partial B}{\partial y}=-2qA,\qquad
%\frac{\partial C}{\partial y}=2rA
%\end{gather}
 We will now describe transition from
the matrix equations (\ref{PM1}) to the 2-component Camassa--Holm
(CH) model. The matrix $P_1$ is parametrized by $q=-2(P_1)_{12}$,
$r=-2(P_1)_{21}$ and $\mathop{\rm Tr}\, (P_1 E)= - (\ln
\tau^0_0)_y$, where $r$ and $q$ are variables of the AKNS model
and $\tau^0_0$ is its tau function. Furthermore,  $A^2+BC=1/16$.
Calculating $\mathop{\rm Tr}\, ((P_1)_y E)$ using the f\/irst
equation of (\ref{PM1}) we get
\begin{gather}
rq = -  (\ln \tau^0_0)_{yy}. \label{eq:rqtau}
\end{gather}
Next, from the second equation of  (\ref{PM1}) we f\/ind
\begin{gather}
q_s = -2 B, \qquad r_s = 2 C \label{eq:rsqs}
\end{gather}
and by taking $\mathop{\rm Tr}\, ((P_1)_s E)$:
\begin{gather}
A = - \frac{1}{2} (\ln \tau^0_0)_{ys}. \label{eq:atau}
\end{gather}
By comparing eqs. (\ref{eq:rqtau}) and (\ref{eq:atau}) we deduce
that $A_y= (rq)_s/2$.

From the third equation of (\ref{PM1}) we derive:
\begin{gather*}
(P_1)_{sy}= \frac{1}{4}  \left(M_0 E M_0^{-1}\right)_y
=\begin{pmatrix}A_y &B_y\\ C_y&-A_y \end{pmatrix} =
\left[\begin{pmatrix}0 &q\\ r&0 \end{pmatrix}, \begin{pmatrix} A
&B\\ C&-A \end{pmatrix}  \right] \label{eq:memy}
\end{gather*}
or in components:
\begin{gather}
B_y =-2 qA, \qquad C_y=2 r A, \qquad A_y=-\frac{1}{2}  (\ln
\tau^0_0)_{syy}= qC-rB \label{eq:bycy}
\end{gather}
and
\begin{gather*}
r_{sy}=4 r A, \qquad q_{sy}=4 q  A. \label{eq:rsy}
\end{gather*}
%The last of expressions in (\ref{eq:rsy}) coincides with the last of
%%expressions
%in (\ref{eq:bycy}) in view of (\ref{eq:atau}).

In view of the fact that the determinant of the matrix $(M_0 E
M_0^{-1})/4$ is equal to the constant, $-1/16$, we choose to
parametrize this matrix in terms of two parameters, $A$ and $f$,
which enter expressions for $B$ and $C$ as follows:
\begin{gather*}
B= e^f \biggl( A-\frac{1}{4} \biggr), \qquad C= -e^{-f} \biggl(
A+\frac{1}{4}\biggr). \label{eq:bcf}
\end{gather*}
Recalling equation (\ref{eq:rsqs}) we easily f\/ind
\begin{gather}
r_s e^f - q_s e^{-f}= 2 (C e^f+Be^{-f})=-1,\qquad r_s e^f + q_s
e^{-f}= 2 (C e^f-Be^{-f})=-4A. \label{eq:rsef}
\end{gather}
Using the f\/irst two identities of equation (\ref{eq:bycy}) and
the fact that
\begin{gather*}
C_ye^f = f_y  \biggl( A+\frac{1}{4} \biggr) -A_y, \qquad B_y
e^{-f} = f_y  \biggl( A-\frac{1}{4} \biggr) +A_y
\end{gather*}
we derive expressions for $r e^f \pm  q e^{-f}$ as follows:
\begin{gather}
r e^f - q e^{-f}= \frac{C_y}{2A} e^f + \frac{B_y}{2A} e^{-f}
=\frac{1}{2A}\biggl(f_y  \biggl( A+\frac{1}{4} \biggr) -A_y +f_y
\biggl( A-\frac{1}{4} \biggr) +A_y \biggr)=f_y \label{eq:refy}
\end{gather}
and
\begin{gather}
r e^f \!+\! q e^{\!-\!f}= \frac{C_y}{2A} e^f \!-\! \frac{B_y}{2A}
e^{\!-\!f} =\frac{1}{2A}\biggl(f_y  \biggl( A\!+\!\frac{1}{4}
\biggr) \!-\!A_y \!-\!f_y  \biggl( A\!-\!\frac{1}{4} \biggr)
\!-\!A_y \biggr) =\frac{f_y}{4A}\!-\!\frac{A_y}{A}.
\label{eq:reAy}
\end{gather}
Taking a derivative of relation $r e^f - q e^{-f}=f_y$ with
respect to variable $s$ we f\/ind, in view of~(\ref{eq:rsef}),
that:
\begin{gather*}
f_{ys} = r_s e^f - q_s e^{-f}+f_s (r e^f + q e^{-f})=-1 +f_s (r
e^f + q e^{-f}) \label{eq:fys}
\end{gather*}
or
\begin{gather}
r e^f + q e^{-f} = 2 g, \label{eq:gpre}
\end{gather}
where we def\/ined:
\begin{gather}
g = \frac{1}{2f_s} (1 +f_{sy}). \label{eq:gdef}
\end{gather}
Comparing two expressions (\ref{eq:reAy}) and (\ref{eq:gpre}) for
the quantity $r e^f + q e^{-f}$ we f\/ind the following relation
\begin{gather}
2 g A =\frac{f_y}{4}-A_y .  %= \frac{1}{f_s} \left( 1 +f_{sy}\right)
\label{eq:impa}
\end{gather}
By adding and subtracting (\ref{eq:gpre}) and (\ref{eq:refy}) we
get
\begin{gather*}
2 r e^f = 2 g +f_y, \qquad 2 q e^{-f} = 2 g-f_y
\end{gather*}
or
\begin{gather*}
q = e^f \biggl( g - \frac{f_y}{2} \biggr), \qquad r = e^{-f}
\biggl( g + \frac{f_y}{2} \biggr). \label{eq:qr-ch}
\end{gather*}
Plugging these expressions into the second relation in eq.
(\ref{eq:rsef}) yields:
\begin{gather*}
-4A= r_s e^f + q_s e^{-f}= \biggl(-f_s r +e^{-f} \biggl( g_s +
\frac{f_{sy}}{2} \biggr) \biggr) e^f+ \biggl(f_s q +e^{f} \biggl(
g_s -\frac{f_{sy}}{2} \biggr) \biggr) e^{-f}
\\ \phantom{-4A}
{}=- f_s (r e^f -q e^{-f}) +2 g_s=- f_s f_y+2 g_s.
\end{gather*}
Thus,
\begin{gather}
A = \frac{1}{4} (f_s f_y-2 g_s)=\frac{1}{4}\biggl(f_s f_y
-\biggl(\frac{1}{f_s}\biggr)_s -\biggl(\frac{f_{sy}}{f_s}\biggr)_s
\biggr). \label{eq:Afsfy}
\end{gather}
Plugging def\/inition (\ref{eq:gdef}) of $g$ into relation
(\ref{eq:impa}) leads to:
%\begin{gather}
%(-g^2-g_y)_s=-f_{sy}f_y-\frac{1}{2} f_s f_{yy}.
%\label{eq:fcond}
%\end{gather}
\begin{gather*}
A =\frac{1}{4} f_s f_y -f_s A_y-A f_{sy}= \frac{1}{4} f_s f_y
-(f_s A)_y. \label{eq:afsfy}
\end{gather*}
Inserting on the left hand side of the above identity the value of
$A$ from equation (\ref{eq:Afsfy}) and multiplying by $-4$ yields
\begin{gather}
\biggl(\frac{1}{f_s}\biggr)_s
=-\biggl(\frac{f_{sy}}{f_s}\biggr)_s+4 (f_s A)_y=
\biggl(-\frac{f_{ss}}{f_s}+4 f_s A\biggr)_y\nonumber
\\ \phantom{\biggl(\frac{1}{f_s}\biggr)_s}{}
=\biggl(-\frac{f_{ss}}{f_s}+f_s^2 f_y
-f_s\biggl(\frac{1+f_{sy}}{f_s}\biggr)_s\biggr)_y =\biggl(f_s^2
f_y - f_{ssy} +\frac{f_{ss}f_{sy}}{f_s} \biggr)_y,
\label{eq:bilcn1}
\end{gather}
where we again used value of $A$ from equation (\ref{eq:Afsfy}).
Note that equation (\ref{eq:bilcn1}) is written solely in terms of
$f$.
%Relation (\ref{eq:bilcn0} can be rewritten as
%\begin{gather}
%\left(f_s^2 f_y - f_{ssy} +\frac{f_{ss}f_{sy}}{f_s} \right)_y =
%- \frac{f_{ss}}{f_s^2}
%\label{eq:bilcn1}
%\end{gather}
For a quantity $u$ def\/ined as:
\begin{gather}
u = f_s^2 f_y - f_{ssy} +\frac{f_{ss}f_{sy}}{f_s} -\frac{1}{2}
\kappa, \label{eq:ufsy}
\end{gather}
with $\kappa$ being an integration constant, it holds from
relation (\ref{eq:bilcn1}) that
\begin{gather}
u_y = \biggl(\frac{1}{f_s}\biggr)_s. \label{eq:uyfs}
\end{gather}
Let us  now denote the product $ f_s^2 f_y$ by $m$. Then from
relations  (\ref{eq:ufsy}) and (\ref{eq:uyfs}) we derive
\begin{gather*}
m = f_s^2 f_y= u \!+\!f_{ssy} \!-\!\frac{f_{ss}f_{sy}}{f_s}
\!+\!\frac{1}{2} \kappa =u \!-\! f_s \biggl(f_s
\biggl(\frac{1}{f_s}\biggr)_s \biggr)_y\!+\!\frac{1}{2} \kappa =u
\!-\!  f_s (f_s  u_y)_y \!+\!\frac{1}{2} \kappa. \label{eq:mdef}
\end{gather*}
%Inserting equations (\ref{eq:ufsy}) and (\ref{eq:uyfs}) we find that $m$
%is equal to
%\begin{gather}
%m=f_s^2 f_y - f_{ssy} +\frac{f_{ss}f_{sy}}{f_s}
%+ f_s \left( \frac{f_{ss}}{f_s}\right)_y = f_s^2 f_y
%\label{eq:mequal}
%\end{gather}
Taking a derivative of $m= f_s^2 f_y $ with respect to $s$ yields
\begin{gather}
m_s = 2 f_y f_s f_{ss}\!+\! f_s^2 f_{sy}=2 m \frac{f_{ss}}{f_s}
\!+\! f_s^2 f_{sy}= \!- 2 m f_s \biggl(\frac{1}{f_s}\biggr)_s
\!+\! f_s^2 f_{sy} = \!-2 m f_s u_y \!+\! f_s^2 f_{sy}.
\label{eq:mseq}
\end{gather}
In terms of the basic quantities $u$ and $\rho=f_s$ of the
two-component Camassa--Holm model equations (\ref{eq:uyfs}) and
(\ref{eq:mseq}) take the following form
\begin{gather}
\rho_s = -\rho^2 u_y , \label{eq:conta}
\\
m_s = -2 m \rho u_y + \rho^2 \rho_{y} \label{eq:ch1}
\end{gather}
for
\begin{gather}
m= u -  \rho  (\rho u_y)_y +\frac{1}{2} \kappa. \label{eq:mdef1}
\end{gather}
Performing an inverse reciprocal transformation $(y,s)\mapsto
(x,t) $ def\/ined by relations:
\begin{gather*}
\label{eq:reciprocal-a} F_x=\rho\,F_y,\qquad
%\pder{}{x} =\rho\,\pder{}{y},\quad
F_t= F_s-\rho\,u\,F_y
%\pder{}{t}=\pder{}{ s}-\rho\,u\,\pder{}{ y}.
\end{gather*}
for an arbitrary function $F$, we f\/ind that equations
(\ref{eq:conta}), (\ref{eq:ch1}) and (\ref{eq:mdef1}) become
\begin{gather}
\rho_t  = - (u \rho)_x , \label{eq:contb}
\\
m_t= -2 m u_x -m_x u+ \rho \rho_x , \label{eq:ch2}
\\
m =u - u_{xx} + \frac{1}{2} \kappa \label{eq:m-def}
\end{gather}
in terms of the $ (x,t) $ variables. Equations
(\ref{eq:contb})--(\ref{eq:m-def}) were introduced by Liu and
Zhang in~\cite {LZ} and are called
 the two-component Camassa--Holm  equations (see also \cite{CH2}).

The relation (\ref{eq:bilcn1}) is equivalent to the following
condition
\begin{gather*}
\frac{f_{ss}}{2 f^3_{s}}+ f_{sy}f_y   +\frac{1}{2} f_s f_{yy} -
\frac{f_{ssyy}}{2f_s} +\frac{f_{ssy} f_{sy}}{2 f_s^2 }
+\frac{f_{ss} f_{syy}}{2 f_s^2 } -\frac{f_{ss} f_{sy}^2}{2 f_s^3
}=0, \label{eq:bilcn0}
\end{gather*}
which f\/irst appeared in \cite{CH2}.

Comparing equations (\ref{eq:Afsfy}) and (\ref{eq:ufsy}) we f\/ind
that
\begin{gather*}
4 f_s A= u+\frac{1}{2} \kappa +\frac{f_{ss}}{f_s}=
u-u_x+\frac{1}{2} \kappa
\end{gather*}
where
\begin{gather*}
4 f_s A= - \frac{r_{ss}}{2 r_s} \left(A+\frac{1}{4}\right)+
\frac{q_{ss}}{2 q_s} \left(A-\frac{1}{4}\right)= -
\frac{r_{ss}q_s}{2\left(A-\frac{1}{4}\right)}+
 \frac{q_{ss}r_s}{2\left(A+\frac{1}{4}\right)}
\end{gather*}
or
\begin{gather*}
4 f_s A=-2 f_s (\ln \tau^0_0)_{sy}=-2  (\ln \tau^0_0)_{sx}.
\end{gather*}
These relations give $u$, $f_s$ in terms of the AKNS quantities
$r$, $q$, $\tau^0_0$.

\section[A second reduction and the Cecotti-Vafa equations]{A second reduction and the Cecotti--Vafa equations}
In order to obtain the Cecotti--Vafa equations we def\/ine an
automorphism of order 4 on the Clif\/ford algebra by
\begin{gather*}
\omega(\psi_{\mathfrak k}^{\pm (j)}) =(-)^{{\mathfrak
k}+\frac{1}{2}}i\psi_{\mathfrak k}^{\mp (j)} \label{omega}
\end{gather*}
then $\omega(\psi^{\pm (j)}(z))=i\psi^{\mp (j)}(-z)$ and
\begin{gather*}
\omega(\alpha_m^{(j)})=-(-)^m\alpha_m^{(j)},\qquad \omega(Q_j^{\pm
1})=iQ_j^{\mp 1}(-)^{\alpha_0^{(j)}}.
\end{gather*}
For the derivation of the last equation see \cite{L}. Next, let
\begin{gather*}
\omega(|0\rangle)=|0\rangle,\qquad\mbox{and }\quad \omega(\langle
0|)=\langle 0|,
\end{gather*}
then this induces also an automorphism on the representation
spaces $F$ and $F^*$. It is straightforward to check that
\begin{gather*}
\omega(|\alpha\rangle)=(-)^{\sum\limits_{j=1}^n
\frac{1}{2}|\alpha_j|(|\alpha_j|-1)} i^{\sum\limits_{j=1}^n
|\alpha_j|}|-\alpha\rangle,
\\
\omega(\langle \alpha|)=(-)^{\sum\limits_{j=1}^n
\frac{1}{2}|\alpha_j|(|\alpha_j|+1)} i^{\sum\limits_{j=1}^n
|\alpha_j|}\langle -\alpha|.
\end{gather*}
Since
\begin{gather*}
\omega(\psi_{\mathfrak k}^{\lambda (j)}\psi_{-\mathfrak
k}^{-\lambda (j)})= \psi_{\mathfrak k}^{-\lambda
(j)}\psi_{-\mathfrak k}^{\lambda (j)}
\end{gather*}
one easily deduces that from (\ref{basis}) and (\ref{basis2}) that
\begin{gather}
\langle\omega(\langle V({\mathfrak r}_1,\ldots, {\mathfrak
r}_m,{\mathfrak s}_1,\ldots, {\mathfrak s}_q)|)|\omega
(|V({\mathfrak i}_1,\ldots, {\mathfrak i}_k,{\mathfrak
j}_1,\ldots, {\mathfrak j}_\ell)\rangle)\rangle\nonumber
\\ \qquad
{}=\langle V({\mathfrak r}_1,\ldots, {\mathfrak r}_m,{\mathfrak
s}_1,\ldots, {\mathfrak s}_q) |V({\mathfrak i}_1,\ldots,
{\mathfrak i}_k,{\mathfrak j}_1,\ldots, {\mathfrak
j}_\ell)\rangle. \label{qqq}
\end{gather}
Assume for the second reduction that our $A$, which commutes with
$\Omega$, also satisf\/ies
\begin{gather}
\omega (A)=A, \label{oma}
\end{gather}
then  one deduces from (\ref{qqq}) that
\begin{gather*}
\omega(\tau_\alpha^\beta (t,u))=\tau_\alpha^\beta (t,u).
\end{gather*}
On the other hand if we use the def\/inition (\ref{tauA}) for the
tau-functions one deduces
\begin{gather*}
\omega( \tau_\alpha^\beta (t_m^{(j)},u_p^{(q)}))
\!=\!(-)^{\sum\limits_{j=1}^n \frac{1}{2}|\alpha_j|(|\alpha_j|+1)+
\frac{1}{2}|\beta_j|(|\beta_j|-1)}i^{\sum\limits_{j=1}^n
|\alpha_j|+|\beta_j|} \tau_{-\alpha}^{-\beta}
(-(-)^mt_m^{(j)},-(-)^pu_p^{(q)}).
\end{gather*}
From now on lets write $\tilde t_m^{(j)}=-(-)^mt_m^{(j)}$,
then combining the above two equations, one has when~(\ref{oma})
holds:
\begin{gather*}
\tau_\alpha^\beta (t,u)= (-)^{\sum\limits_{j=1}^n
\frac{1}{2}|\alpha_j|(|\alpha_j|+1)+\frac{1}{2}|\beta_j|(|\beta_j|-1)}
i^{\sum\limits_{j=1}^n
|\alpha_j|+|\beta_j|}\tau_{-\alpha}^{-\beta} (\tilde t, \tilde u).
\label{qqqq}
\end{gather*}
In particular
\begin{gather*}
\tau_0^0 \left(t,u\right)=\tau_0^0(\tilde t,\tilde u),\qquad
\tau_{\epsilon_k-\epsilon_\ell}^0 (t,u)=
-\tau_{\epsilon_\ell-\epsilon_k}^0(\tilde t,\tilde u),\qquad
\tau_{\epsilon_k}^{\epsilon_\ell} (t,u)=
\tau_{-\epsilon_\ell}^{-\epsilon_k}(\tilde t,\tilde u).
\label{qqqqq}
\end{gather*}
Now substituting this in (\ref{wwave}) for $\alpha=\beta=0$ we
obtain that
\begin{gather*}
\psi^{+(0)}(0,0,x,t,u,z)=\psi^{-(0)}(0,0,x,\tilde t,\tilde u,-z),
\\
\psi^{+(1)}(0,0,x,t,u,z)=-\psi^{-(1)}(0,0,x,\tilde t,\tilde u,-z).
\end{gather*}

{\it Assume from now on that $A$ satisfies \eqref{A} and
\eqref{oma}, viz. that both reductions hold}, then from
\eqref{third}, \eqref{invP1}, \eqref{invP11} we deduce that
\begin{gather*}
P^{+(0)}(0,0,x,t,u,z)^{-1}=P^{+(0)}(0,0,x,\tilde t, \tilde
u,-z)^{T},
\\
P^{+(1)}(0,0,x,t,u,z)^{-1}=-P^{+(1)}(0,0,x,\tilde t, \tilde
u,-z)^{T}z^2,
\end{gather*}
then using the def\/inition (\ref{loop0}) one f\/inally obtains
\begin{gather*}
P^{(0)}(0,0,x,t,u,z)^{-1}=P^{(0)}(0,0,x,\tilde t, \tilde u,-z)^{T},\\
P^{(1)}(0,0,x,t,u,z)^{-1}=P^{(1)}(0,0,x,\tilde t, \tilde
u,-z)^{T}.
\end{gather*}
Now putting all $t_{2m}^{(j)}= u_{2m}^{(j)}=0$, this gives the
following equations for $P_1$, $M_0$, $M_1$ and $\bar P_1$:
\begin{gather}
P_1^T=P_1,\qquad M_0^T=M_0^{-1},\qquad M_1^T=M_0^T M_1M_0^T,\qquad
\bar P_1^T=\bar P_1. \label{transp}
\end{gather}
Now denote
\begin{gather*}
P_1=(\beta_{ij})_{1\le i,j\le n},\qquad M_0=(m_{ij})_{1\le i,j\le
n} \qquad \bar P_1=(\bar\beta_{ij})_{1\le i,j\le n},
\end{gather*}
then the equations (\ref{transp}) and (\ref{PM2}) give the
following system of Cecotti--Vafa equations (see
e.g.~\cite{AGZ,CV,D}):
\begin{gather}
\beta_{ij}=\beta_{ji},\quad\bar\beta_{ij}=\bar\beta_{ji},\quad
\partial_j\beta_{ik}=\beta_{ij}\beta_{jk},\quad
\partial_{-j}\bar\beta_{ik}=\bar\beta_{ij}\bar\beta_{jk},\qquad i,j,k\quad {\rm distinct},\nonumber
\\
\sum_{j=1}^n\partial_j\beta_{ik}=\sum_{j=1}^n\partial_j\bar\beta_{ik}=
\sum_{j=1}^n\partial_{-j}\beta_{ik}=\sum_{j=1}^n\partial_{-j}\bar\beta_{ik}=0,\qquad
i\ne k,\nonumber
\\
\partial_{-j}\beta_{ik}=m_{ij}m_{kj},\qquad
\partial_{j}\bar\beta_{ik}=m_{ji}m_{jk}\qquad i\ne k,\nonumber
\\
\sum_{j=1}^n\partial_jm_{ik}=\sum_{j=1}^n\partial_{-j}m_{ik}=0,\qquad
\partial_j m_{ik}=\beta_{ij}m_{jk},\qquad\partial_{-j}
m_{ik}=m_{ij}\bar\beta_{jk}. \label{DE}
\end{gather}

\section{Homogeneity}
Sometimes one wants to obtain solutions of the Cecotti--Vafa
equations that satisfy certain homogeneity condition (see e.g.~\cite{D}). 
For this we introduce the $L_0$ element of a Virasoro
algebra. The most natural def\/inition  in our construction of the
Clif\/ford algebra is the one given in terms of the oscillator
algebra.
\begin{gather}
L_0=\sum_{j=1}^{n} \frac{1}{2}(\alpha_0^{(j)})^2+\sum_{k=1}^\infty
\alpha_{-k}^{(j)}\alpha_{k}^{(j)}. \label{h1}
\end{gather}
It is straightforward to check that
\begin{gather*}
[L_0,\alpha_k^{(j)}]=-k\alpha_k^{(j)}
\end{gather*}
and
\begin{gather*}
\langle \beta|L_0=\frac{1}{2}\sum_{j=1}^n |\beta_j|^2 \langle
\beta |, \qquad L_0| \beta\rangle=\frac{1}{2}\sum_{j=1}^n
|\beta_j|^2 | \beta\rangle.
\end{gather*}
Moreover, one also has
\begin{gather*}
\big[L_0,\psi_{\mathfrak k}^{\pm(j)}\big]=-{\mathfrak
k}\psi_{\mathfrak k}^{\pm(j)}.
\end{gather*}
Assume from now on that our operator $A$ that commutes with
$\Omega$ is homogeneous of degree $p$ with respect to $L_0$, i.e.,
\begin{gather*}
[L_0, A]=pA. \label{hom}
\end{gather*}
We then calculate
\begin{gather*}
\langle \alpha|L_0\tilde A|\beta\rangle= \langle
\alpha|L_0\prod_{j=1}^n \Gamma_+^{(j)}(t^{(j)}) A\prod_{k=1}^n
\Gamma_-^{(k)}(-u^{(k)}) |\beta\rangle. \label{h2}
\end{gather*}
It is straightforward to see that this is equal to
\begin{gather*}
\frac{1}{2}\sum_{j=1}^n |\alpha_j|^2 \langle \alpha|L_0\tilde
A|\beta\rangle.
\end{gather*}
On the other hand using
\begin{gather*}
[L_0,\Gamma_+^{(j)}(t^{(j)})]=  -\sum_{k=1}^\infty
kt_k^{(j)}\alpha_k^{(j)}\Gamma_+^{(j)}(t^{(j)})
=-\sum_{k=1}^\infty kt_k^{(j)}\frac{\partial}{\partial
t_k^{(j)}}\Gamma_+^{(j)}(t^{(j)})
\end{gather*}
and
\begin{gather*}
[L_0,\Gamma_-^{(j)}(-u^{(j)}]= \sum_{k=1}^\infty
ku_k^{(j)}\frac{\partial}{\partial u_k^{(j)}}
\Gamma_-^{(j)}(-u^{(j)})
\end{gather*}
we also have:
\begin{gather*}
\langle \alpha|L_0\tilde A|\beta\rangle=
\Biggl(p+\frac{1}{2}\sum_{j=1}^n |\beta_j|^2+ \sum_{j=1}^n
\sum_{k=1}^\infty ku_k^{(j)}\frac{\partial}{\partial u_k^{(j)}}-
kt_k^{(j)}\frac{\partial}{\partial t_k^{(j)}}\Biggr) \langle
\alpha|\tilde A|\beta\rangle.
\end{gather*}
Now let
\begin{gather*}
{\cal E}=\sum_{k=1}^\infty kt_k^{(j)}\frac{\partial} {\partial
t_k^{(j)}}- ku_k^{(j)}\frac{\partial}{\partial u_k^{(j)}},
\label{h3}
\end{gather*}
then
\begin{gather*}
{\cal E}\tau_\alpha^\beta(t,u)= \Biggl( p+\frac{1}{2}\sum_{j=1}^n
|\beta_j|^2-|\alpha_j|^2\Biggr) \tau_\alpha^\beta(t,u), \label{h4}
\end{gather*}
in particular
\begin{gather*}
{\cal E}\tau_0^0(t,u)=p\tau_0^0(t,u),\qquad {\cal
E}\tau_{\epsilon_k-\epsilon_\ell}^0(t,u)=(p-1)
\tau_{\epsilon_k-\epsilon_\ell}^0(t,u),\qquad {\cal
E}\tau_{\epsilon_k}^{\epsilon_\ell}(t,u)=
p\tau_{\epsilon_k}^{\epsilon_\ell}(t,u). \label{h5}
\end{gather*}
Assume now that we also have imposed the f\/irst and second
reduction, then one has
\begin{gather*}
{\cal E}P_1=-P_1,\qquad {\cal E}P_0=0,\qquad {\cal E}M_1=M_1
\end{gather*}
and thus also
\begin{gather*}
{\cal E}\bar P_1=\bar P_1.
\end{gather*}
Putting all $t_m^{(j)}=u_m^{(j)}=0$ for all $m>1$ and all $1\le
j\le n$ one has that the $\beta_{ij}$, $m_{ij}$ and
$\bar\beta_{ij}$ not only satisfy (\ref{DE}), but also
\begin{gather}
\sum_{j=1}^n
(t^j\partial_j-u^j\partial_{-j})\beta_{ij}=-\beta_{ij}, \qquad
\sum_{j=1}^n ( t^j\partial_j-u^j\partial_{-j})m_{ij}=0,\nonumber
\\
\sum_{j=1}^n
(t^j\partial_j-u^j\partial_{-j})\bar\beta_{ij}=\bar\beta_{ij},
\label{h6}
\end{gather}
for $t^j=t_1^{(j)}$ and $u^j=u_1^{(j)}$.

Note that in Section 10 we will construct explicit solutions of
(\ref{DE}). These solutions are however not homogeneous, so they
do not satisfy (\ref{h6}). We will construct such solutions in
a~forthcoming publication.

\section{Explicit construction of solutions in the AKNS case}
We will construct an operator $A$ that satisf\/ies (\ref{commOm})
and (\ref{A}) in the Camassa--Holm case, i.e., the case that
$n=2$. Now using Proposition \ref{prop2.1}, we see that the
element
\begin{gather*}
A_k=(a_1^{(1)}\psi^{\lambda_1
(1)}(z_1)+a_1^{(2)}\psi^{\lambda_1(2)}(z_1))
(a_2^{(1)}\psi^{\lambda_2
(1)}(z_2)+a_2^{(2)}\psi^{\lambda_2(2)}(z_2))\cdots
\\ \phantom{A_k=}
{}\cdots(a_{k}^{(1)}\psi^{\lambda_{k}
(1)}(z_{k})+a_{k}^{(2)}\psi^{\lambda_{k}(2)}(z_{k})), \label{CH1}
\end{gather*}
satisf\/ies condition (\ref{commOm}). Using (\ref{Q}) and the
def\/inition of the fermionic f\/ields, we see that
\begin{gather*}
Q_1Q_2 A_k=z_1^{\lambda_1}z_2^{\lambda_2}\cdots
z_{k}^{\lambda_{k}}A_kQ_1Q_2.
\end{gather*}
Thus $A_k$ also satisf\/ies (\ref{A}). Since we want $\tau^0_0\ne
0$, we take $A_{2k}$ and will assume that
\begin{gather*}
\lambda_1=\lambda_2=\cdots=\lambda_k=+ \qquad \mbox{and}\qquad
\lambda_{k+1}=\lambda_{k+2}=\cdots=\lambda_{2k}=-.
\end{gather*}
We now want to calculate $\tau_{\epsilon_i-\epsilon_j}^0$ and
$\tau_{\epsilon_i}^{\epsilon_j}$, for $1\le i,j\le 2$ in order to
get some solutions related to the Camassa--Holm equation. Let us
start with $\tau_0^0$:
\begin{gather*}
\tau_0^0(t,u)=\langle 0| \prod_{i=1}^2 \Gamma_+^{(i)}(t^{(i)}) \,
A_{2k} \, \prod_{j=1}^2\Gamma_-^{(j)}(-u^{(j)}) |0\rangle
\\ \phantom{\tau_0^0(t,u)}
{}=\sum_{\ell_1+\cdots \ell_k=\ell_{k+1}+\cdots\ell_{2k}}
a_1^{(\ell_1)}a_2^{(\ell_2)}\cdots a_{2k}^{(\ell_{2k})} \langle 0|
\prod_{i=1}^2 \Gamma_+^{(i)}(t^{(i)}) \,
\psi^{+(\ell_1)}(z_1)\psi^{+(\ell_2)}(z_2)\cdots
\\ \phantom{\tau_0^0(t,u)=}{}
\cdots\psi^{+(\ell_k)}(z_k)
\psi^{-(\ell_{k+1})}(z_{k+1})\cdots\psi^{-(\ell_{2k})}(z_{2k})
\prod_{j=1}^2\Gamma_-^{(j)}(-u^{(j)}) |0\rangle
\\ \phantom{\tau_0^0(t,u)}{}
=\gamma(t^{(1)},-u^{(1)})\gamma(t^{(2)},-u^{(2)})
\sum_{\ell_1+\cdots \ell_k=\ell_{k+1}+\cdots\ell_{2k}}
\prod_{i=1}^k a_i^{(\ell_i)}a_{i+k}^{(\ell_{i+k})}
\gamma(t^{(\ell_i)},[z_i])
\\ \phantom{\tau_0^0(t,u)=}{}
\times\gamma(-t^{(\ell_{k+i})},[z_{k+i}])\gamma(u^{(\ell_i)},[z_i^{-1}])
\gamma(-u^{(\ell_{k+i})},[z_{k+i}^{-1}])
\\ \phantom{\tau_0^0(t,u)=}{}\times
\langle 0|\psi^{+(\ell_1)}(z_1)\psi^{+(\ell_2)}(z_2)\cdots
\cdots\psi^{+(\ell_k)}(z_k)
\psi^{-(\ell_{k+1})}(z_{k+1})\cdots\psi^{-(\ell_{2k})}(z_{2k})|0\rangle.
\end{gather*}
So we have to calculate explicitly:
\begin{gather*}
\langle
0|\psi^{+(\ell_1)}(z_1)\psi^{+(\ell_2)}(z_2)\cdots\psi^{+(\ell_k)}(z_k)
\psi^{-(\ell_{k+1})}(z_{k+1})\cdots\psi^{-(\ell_{2k})}(z_{2k})|0\rangle
\\ \qquad
{}=\sigma(\ell_1,\ell_2,\ldots \ell_{2k})\frac{\prod\limits_{1\le
i<j\le k}
(z_i-z_j)^{\delta_{\ell_i,\ell_j}}(z_{k+i}-z_{k+j})^{\delta_{\ell_{k+i},\ell_{k+j}}}}
{\prod\limits_{1\le i,j\le
k}(z_i-z_{k+j})^{\delta_{\ell_i,\ell_{k+j}}}},
\end{gather*}
where
\begin{gather*}
\sigma(\ell_1,\ell_2,\ldots,\ell_{2k})=\langle
0|Q_{\ell_1}Q_{\ell_2}\cdots Q_{\ell_k}Q_{\ell_{k+1}}^{-1}\cdots
Q_{\ell_{2k}}^{-1}|0\rangle.
\end{gather*}
Note that the above expression takes only values $-1,\ 0$ or $1$.
Thus
\begin{gather*}
\tau_0^0(t,u)=\gamma(t^{(1)},-u^{(1)})\gamma(t^{(2)},-u^{(2)})
\\ \qquad
{}\times\sum_{\ell_1+\cdots \ell_k=\ell_{k+1}+\cdots\ell_{2k}}
\sigma(\ell_1,\ell_2,\ldots,\ell_{2k}) \frac{\prod\limits_{1\le
i<j\le k}(z_i-z_j)^{\delta_{\ell_i,\ell_j}} (z_{k+i}-z_{k+j}
)^{\delta_{\ell_{k+i},\ell_{k+j}}}} {\prod\limits_{1\le i,j\le
k}(z_i-z_{k+j})^{\delta_{\ell_i,\ell_{k+j}}}}
\\ \qquad
{}\times\prod_{i=1}^k  a_i^{(\ell_i)}a_{i+k}^{(\ell_{i+k})}
\gamma(t^{(\ell_i)},[z_i])\gamma(-t^{(\ell_{k+i})},[z_{k+i}])
\gamma(u^{(\ell_i)},[z_i^{-1}])\gamma(-u^{(\ell_{k+i})},[z_{k+i}^{-1}]).
\end{gather*}
Now $\tau_{\epsilon_i-\epsilon_j}^0$ for $i\ne j$ has the same
expression, except that we have to take the summation over
\begin{gather*}
j+\ell_1+\ell_2+\cdots+\ell_k=
i+\ell_{k+1}+\ell_{k+2}+\cdots+\ell_{2k}
\end{gather*}
and replace the $\sigma$ by
\begin{gather*}
(-)^i\sigma(j,\ell_1,\ell_2,\ldots,\ell_{2k},i).
\end{gather*}
For $\tau_{\epsilon_i}^{\epsilon_j}$, where $i$ can be equal to
$j$, we f\/ind
\begin{gather*}
\tau_{\epsilon_i}^{\epsilon_j}(t,u)=(-)^{1-\delta_{ij}}
\gamma(t^{(1)},-u^{(1)})\gamma(t^{(2)},-u^{(2)})
\sum_{j+\ell_1+\cdots \ell_k=i+\ell_{k+1}+\cdots\ell_{2k}}
\sigma(j,\ell_1,\ell_2,\ldots,\ell_{2k},i)
\\ \qquad
{}\times z_1^{\delta_{j\ell_1}}z_2^{\delta_{j\ell_2}}\cdots
z_k^{\delta_{j\ell_k}}z_{k+1}^{-\delta_{j\ell_{k+1}}}\cdots
z_{2k}^{-\delta_{j\ell_{2k}}} \frac{\prod\limits_{1\le i<j\le
k}(z_i-z_j)^{\delta_{\ell_i,\ell_j}} (z_{k+i}-z_{k+j}
)^{\delta_{\ell_{k+i},\ell_{k+j}}}} {\prod\limits_{1\le i,j\le
k}(z_i-z_{k+j})^{\delta_{\ell_i,\ell_{k+j}}}}
\\ \qquad
{}\times\prod_{i=1}^k  a_i^{(\ell_i)}a_{i+k}^{(\ell_{i+k})}
\gamma(t^{(\ell_i)},[z_i])\gamma(-t^{(\ell_{k+i})},[z_{k+i}])
\gamma(u^{(\ell_i)},[z_i^{-1}])\gamma(-u^{(\ell_{k+i})},[z_{k+i}^{-1}]).
\end{gather*}
We now give the simplest of such expressions. We take $k=1$ and
put all $t_j^{(i)}=u_j^{(i)}=0$ for $j>2$, then
\begin{gather*}
\tau_0^0(t,u)=\frac{T(t,u)}{z_1-z_2}(T_{11}(t,u)+T_{22}(t,u)),\qquad
\tau_{\epsilon_i-\epsilon_j}^0(t,u)=(-)^{i+1}T(t,u)T_{ij}(t,u),\nonumber
\\
\tau_{\epsilon_i}^{\epsilon_j}(t,u)=z_2^{-1}T(t,u)T_{ij}(t,u),\qquad\mbox{for}\qquad
i\ne j,\nonumber
\\
\tau_{\epsilon_i}^{\epsilon_i}(t,u)=\frac{T(t,u)}{z_1-z_2}
\biggl(\biggl(\frac{z_1}{z_2}\biggr)^{\delta_{1i}}T_{11}(t,u)
+\biggl(\frac{z_1}{z_2}\biggr)^{\delta_{2i}}T_{22}(t,u)\biggr),
\label{simp}
\end{gather*}
where
\begin{gather}
T(t,u)= e^{-t_1^{(1)} u_1^{(1)}-t_1^{(2)} u_1^{(2)}-2t_2^{(1)}
u_1^{(2)}-2t_2^{(2)} u_2^{(2)}},\nonumber
\\
T_{ij}(t,u)=T_{ij}(t,u, z_1,z_2)\nonumber
\\ \phantom{T_{ij}(t,u)}
{}=a_1^{(i)}a_2^{(j)}e^{t_1^{(i)}z_1+ u_1^{(i)}z_1^{-1}
+t_2^{(i)}z_1^2+ u_2^{(i)}z_1^{-2}-(t_1^{(j)}z_2+u_1^{(j)}z_2^{-1}
+t_2^{(j)}z_2^{2}+ u_2^{(j)}z_2^{-2}}. \label{TT}
\end{gather}
In order to determine $M_0$, $M_1$ and $P_1$, see (\ref{ttau}), we
also calculate
\begin{gather*}
\partial_j\tau_0^0(t,u)=-u_1^{(j)}
\frac{T(t,u)}{z_1-z_2}(T_{11}(t,u)+T_{22}(t,u))+T(t,u)T_{jj}(t,u),\nonumber
\\
\partial_{-j}\tau_{\epsilon_i}^{\epsilon_j}(t,u)=
-(z_2^{-1}t_1^{(j)}+z_2^{-2})T(t,u)T_{ij}(t,u),\qquad\mbox{for}\qquad
i\ne j,\nonumber
\\
\partial_{-i}\tau_{\epsilon_i}^{\epsilon_i}(t,u)=-t_1^{(i)}
\frac{T(t,u)}{z_1-z_2}
\biggl(\biggl(\frac{z_1}{z_2}\biggr)^{\delta_{1i}}T_{11}(t,u)
+\biggl(\frac{z_1}{z_2}\biggr)^{\delta_{2i}}T_{22}(t,u)\biggr)
-z_2^{-2}T(t,u)T_{ii}(t,u). \label{simp1}
\end{gather*}
Now make the change of variables (\ref{ys}) and choose
$t_2^{(i)}=u_2^{(i)}=0$. One thus obtains
\begin{gather}
t_1^{(1)}=\bar y+y,\qquad t_1^{(2)}=\bar y-y,\qquad
u_1^{(1)}=\frac{\bar s+s}4,\qquad u_1^{(2)}=\frac{\bar s-s}4
\label{change}
\end{gather}
and
\begin{gather}
T(y,s)=e^{-\frac{1}{4}(\bar y+y)(\bar s+s)-\frac{1}{4}(\bar
y-y)(\bar s-s)},\nonumber
\\
T_{ij}(y,s,z_1,z_2)\!=\!T_{ij}(y,s)\!=\!a_1^{(i)}a_2^{(j)}
e^{(\bar y-(-)^iy)z_1\!+\! \frac{1}{4}(\bar s-(-)^is)z_1^{-1}
\!-\!((\bar y-(-)^jy)z_2+ \frac{1}{4}(\bar s-(-)^js)z_2^{-1})}\!.
\label{TTT}
\end{gather}
Note that in the calculation of $M_0$, $M_1$ and $P_1$, see
(\ref{ttau}), $T(y,s)$ drops out. Thus
\begin{gather*}
M_0=\biggl(\delta_{ij}+\frac{(z_1z_2^{-1}-1)T_{ij}(y,s)}{T_{11}(y,s)
+T_{22}(y,s)}\biggr)_{1\le i,j\le 2},
\\
M_1=\biggl(\delta_{ij}(\bar y-(-)^j
y)+\frac{(z_1z_2^{-1}-1)(z_2^{-1} +\bar y-(-)^j y)
T_{ij}(y,s)}{T_{11}(y,s)+T_{22}(y,s)}\biggr)_{1\le i,j\le 2},
\\
P_1=\biggl(\delta_{ij}\frac{\bar s-(-)^j
s}4-\frac{(z_1-z_2)T_{ij}(y,s)}{T_{11}(y,s)
+T_{22}(y,s)}\biggr)_{1\le i,j\le 2}.
\end{gather*}
One has  $\det M_0={z_1}/{z_2}$ and $M_0$, $M_1$ and $P_1$ satisfy
(\ref{PM1}). The above solutions should in principle be closely
related to the ones obtained  in \cite{Aratyn:2005pg} using vertex
method. Finally, we give the Camassa--Holm function $f$ for this
case:
\begin{gather*}
f(y,s)=\log  \frac{a_1^{(1)}a_2^{(1)}z_1e^{\frac{s}{2z_1}+2yz_1}
+a_1^{(2)}a_2^{(2)}z_2e^{\frac{s}{2z_2}+2yz_2} } {
(z_2-z_1)a_1^{(2)}a_2^{(1)} }.
\end{gather*}

We give expression for tau  functions for $k=2$. Again we set
$t_j^{(i)}=u_j^{(i)}=0$ for $i,j>2$ and use $T (t,u)$ and $T_{ij}
(t,u,z_k,z_l)$ with $i,j=1,2$, $(k,\ell)=(1,3)$, $(k,\ell)=(2,4)$.
For brevity we will denote
\begin{gather*}
T=T(t,u),\qquad T_{ij}^{k,\ell}=T_{ij} (t,u,z_k,z_\ell).
\end{gather*}
Then, the $\tau^0_0$ function is given by
\begin{gather*}
\tau^0_0= T  \biggl[\frac{(z_1-z_2)(z_3-z_4)}{(z_1-z_3)
(z_2-z_3)(z_1-z_4)(z_2-z_4)} (T_{11}^{1,3} T_{11}^{2,4}
+T_{22}^{1,3} T_{22}^{2,4})
\\ \qquad
{}- \frac{1}{(z_1-z_3)(z_2-z_4)}(T_{11}^{1,3} T_{22}^{2,4}
+T_{22}^{1,3} T_{11}^{2,4})
+\frac{1}{(z_1-z_4)(z_2-z_3)}(T_{12}^{1,3} T_{21}^{2,4}
+T_{21}^{1,3} T_{12}^{2,4}) \biggr].
\end{gather*}
In the following formulas for remaining tau functions we choose
indices $i,j=1,2$ so that $i\ne j$. Then
\begin{gather*}
\tau_{\epsilon_i-\epsilon_j}= (-1)^{i+1}T
\biggl[\frac{(z_1-z_2)}{(z_1-z_3) (z_2-z_3)}T_{ii}^{1,3}
T_{ij}^{2,4} -\frac{(z_1-z_2)}{(z_1-z_4) (z_2-z_4)}T_{ij}^{1,3}
T_{ii}^{2,4}
\\ \phantom{\tau_{\epsilon_i-\epsilon_j}=}
{} + \frac{(z_3-z_4)}{(z_2-z_3)(z_2-z_4)} T_{ij}^{1,3}
T_{jj}^{2,4} -\frac{(z_3-z_4)}{(z_1-z_3)(z_1-z_4)} T_{jj}^{1,3}
T_{ij}^{2,4}  \biggr],
\\
\tau_{\epsilon_j}^{\epsilon_i}= -T \biggl[\frac{z_1(z_3-z_4)}{z_3
z_4(z_1-z_3) (z_1-z_4)}T_{ii}^{1,3} T_{ji}^{2,4}
-\frac{z_2(z_3-z_4)}{z_3z_4 (z_2-z_3) (z_2-z_4)}T_{ji}^{1,3}
T_{ii}^{2,4}
\\ \phantom{\tau_{\epsilon_j}^{\epsilon_i}=}
{}+ \frac{(z_1-z_2)}{z_3(z_1-z_4)(z_2-z_4)} T_{ji}^{1,3}
T_{jj}^{2,4} -\frac{(z_1-z_2)}{z_4(z_1-z_3)(z_2-z_3)} T_{jj}^{1,3}
T_{ji}^{2,4}  \biggr],
\\
\tau_{\epsilon_i}^{\epsilon_i}= T
\biggl[\frac{(z_1-z_2)(z_3-z_4)}{(z_1-z_3)
(z_1-z_4)(z_2-z_3)(z_2-z_4)}\biggl(\frac{z_1z_2}{z_3z_4}
T_{ii}^{1,3} T_{ii}^{2,4} + T_{jj}^{1,3} T_{jj}^{2,4}\biggr)
\\ \phantom{\tau_{\epsilon_i}^{\epsilon_i}=}
{}- \frac{1}{(z_1-z_3)(z_2-z_4)} \biggl( \frac{z_1}{z_3}
T_{ii}^{1,3} T_{jj}^{2,4} +\frac{z_2}{z_4} T_{jj}^{1,3}
T_{ii}^{2,4}\biggr)
\\ \phantom{\tau_{\epsilon_i}^{\epsilon_i}=}
{}+\frac{1}{(z_1-z_4)(z_2-z_3)}\biggl( \frac{z_1}{z_4}T_{ij}^{1,3}
 T_{ji}^{2,4}+\frac{z_2}{z_3}T_{ji}^{1,3}T_{ij}^{2,4}\biggr)\biggr].
\end{gather*}
If we make the change of variables (\ref{change}) then the
formula's (\ref{TT}) change into (\ref{TTT}) and the same
formula's hold for the tau functions.

Another way to construct solutions is as follows. We take an
arbitrary element of the loop algebra of $sl_2(\mathbb{C})$. Such
an element is given by
\begin{gather*}
g=\mathop{\rm Res}_z
a(z)(\psi^{+(1)}(z)\psi^{-(1)}(z)-\psi^{+(2)}(z)\psi^{-(2)}(z)) +
b(z)\psi^{+(1)}(z)\psi^{-(2)}(z)
\\ \phantom{g=}
{}+c(z)\psi^{+(2)}(z)\psi^{-(2)}(z),
\end{gather*}
where $a(z),\ b(z)$ and $c(z)$ are arbitrary functions. Then the
element $A=e^g$ commutes with the action of $\Omega$ and
satisf\/ies (\ref{A}). Hence the corresponding tau-functions will
satisfy the equations of the AKNS hierarchy.

\section[Explicit construction of solutions in the Cecotti-Vafa case]{Explicit construction of solutions in the Cecotti--Vafa case}

We will construct an operator $A$ that satisf\/ies (\ref{commOm}),
(\ref{A}) and (\ref{oma}). We generalize the construction of the
previous section. Using Proposition \ref{prop2.1}, we see that the
element
\begin{gather*}
A_k=\Biggl(\sum_{j_1=1}^n a_1^{(j_1)}\psi^{+
(j_1)}(z_1)\Biggr)\nonumber \Biggl(\sum_{j_1=1}^n
a_1^{(j_1)}\psi^{-(j_1)}(-z_1)\Biggr) \Biggl(\sum_{j_1=1}^n
a_2^{(j_2)}\psi^{+ (j_2)}(z_2)\Biggr)
\\ \phantom{A_k=}
{}\times\Biggl(\sum_{j_1=1}^n a_2^{(j_2)}\psi^{-
(j_2)}(-z_2)\Biggr) \cdots \Biggl(\sum_{j_k=1}^n
a_{k}^{(j_k)}\psi^{+ (j_k)}(z_{k})\Biggr) \Biggl(\sum_{j_k=1}^n
a_{k}^{(j_k)}\psi^{- (j_k)}(-z_{k})\Biggr), \label{C-V1}
\end{gather*}
satisf\/ies condition (\ref{commOm}). Moreover, if we assume  that
\begin{gather}
\sum_{j=1}^n (a_\ell^{(j)})^2=0 \qquad \mbox{for all}\quad 1\le
\ell\le k, \label{C-V2}
\end{gather}
then
\begin{gather*}
\Biggl(\sum_{j=1}^n a_\ell^{(j)}\psi^{+(j)}(z)\Biggr)
\Biggl(\sum_{k=1}^n a_\ell^{(k)}\psi^{-(k)}(y)\Biggr)=-
\Biggl(\sum_{k=1}^n a_\ell^{(k)}\psi^{-(k)}(y)\Biggr)
\Biggl(\sum_{j=1}^n a_\ell^{(j)}\psi^{+(j)}(z)\Biggr).
\end{gather*}
Thus
\begin{gather*}
\omega\Biggl(\!\Biggl(\sum_{j=1}^n
a_\ell^{(j)}\psi^{+(j)}(z)\Biggr)\! \Biggl(\sum_{k=1}^n
a_\ell^{(k)}\psi^{-(k)}(-z)\Biggr)\!\Biggr)= -\Biggl(\sum_{j=1}^n
a_\ell^{(j)}\psi^{-(j)}(-z)\Biggr) \Biggl(\sum_{k=1}^n
a_\ell^{(k)}\psi^{+(k)}(z)\Biggr)
\\ \qquad
{}=\Biggl(\sum_{k=1}^n a_\ell^{(k)}\psi^{+(k)}(z)\Biggr)
\Biggl(\sum_{j=1}^n a_\ell^{(j)}\psi^{-(j)}(-z)\Biggr)
\end{gather*}
and hence $A_k$ also satisf\/ies (\ref{oma}). Let us calculate for
$k=1$ the corresponding tau functions. One f\/inds
\begin{gather*}
\tau_0^0=\frac{T}{2z}\sum_{j=1}^n T_{jj}(z),\qquad
\tau_{\epsilon_i-\epsilon_j}^0=(-)^{|\epsilon_j|_i}T(z)T_{ij}(z),\qquad
\tau_{\epsilon_i}^{\epsilon_j}=-\frac{T}{z}T_{ij}(z)\qquad
\mbox{for}\qquad
i\ne j,\\
\tau_{\epsilon_i}^{\epsilon_i}=\frac{T}{2z}\sum_{j=1}^n
(-)^{\delta_{ij}}T_{jj}(z),
\end{gather*}
where
\begin{gather*}
T=\prod_{i=1}^n \gamma(t^{(i)},-u^{(i)}),\nonumber
\\
T_{ij}(z_k)=a_k^{(i)}a_k^{(j)}\gamma(t^{(i)},[z_k])\gamma(-t^{(j)},[-z_k])
\gamma(u^{(i)},[z_k^{-1}])\gamma(-u^{(j)},[-z_k^{-1})]).
\label{Ts}
\end{gather*}

{\samepage If we only keep $t_1^{(i)}$ and $u_1^{(i)}$ and put all
higher times equal to zero, we f\/ind that
\begin{gather*}
\beta_{ij}=z m_{ij}=z^2\bar\beta_{ij},\qquad\mbox{for}\qquad i\ne
j
\end{gather*}
and that
\begin{gather*}
m_{ij}=\delta_{ij}-\frac{2a_1^{(i)}a_1^{(j)}e^{(t_1^{(i)}
+t_1^{(j)})z+(u_1^{(i)}+u_1^{(j)})z^{-1}}}
{\sum\limits_{i=1}^n(a_1^{(i)})^2e^{2t_1^{(i)}z+2u_1^{(i)}z^{-1}}},
\end{gather*}
where (\ref{C-V2}) still holds. These $\beta_{ij}$, $m_{ij}$ and
$\bar\beta_{ij}$ satisfy (\ref{DE}).

}

For the case that $k=2$ we describe the tau functions
\begin{gather*}
\tau_0^0=T\biggl(\frac{(z_1-z_2)^2}{4z_1z_2(z_1+z_2)^2}\sum_{i=1}^n
T_{ii}(z_1)T_{ii}(z_2)+\sum_{i\ne
j}\frac{1}{4z_1z_2}T_{ii}(z_1)T_{jj}(z_2)
\\ \phantom{\tau_0^0=}
{} -\sum_{i\ne
j}\frac{1}{(z_1+z_2)^2}T_{ij}(z_1)T_{ji}(z_2)\biggr),
\\
\tau_{\epsilon_k}^{\epsilon_k}=T\biggl(\frac{(z_1-z_2)^2}{4z_1z_2(z_1+z_2)^2}
\sum_{i=1}^n T_{ii}(z_1)T_{ii}(z_2)+\sum_{i\ne
j}(-)^{\delta_{ik}+\delta_{jk}}
\frac{1}{4z_1z_2}T_{ii}(z_1)T_{jj}(z_2)
\\ \phantom{\tau_{\epsilon_k}^{\epsilon_k}=}
{}-\sum_{i\ne j}\biggl(-\frac{z_1}{z_2}\biggr)^{\delta_{ik}}
\biggl(-\frac{z_2}{z_1}\biggr)^{\delta_{jk}}
\frac{1}{(z_1+z_2)^2}T_{ij}(z_1)T_{ji}(z_2)\biggr),
\\
\tau_{\epsilon_i-\epsilon_j}^0=T\biggl(\frac{z_2-z_1}{2(z_1+z_2)}
\biggl(\frac{1}{z_2}T_{ij}(z_1)(T_{ii}(z_2)+T_{jj}(z_2))
-\frac{1}{z_1}(T_{jj}(z_1)+T_{ii}(z_1))T_{ij}(z_2)\biggr)
\\ \phantom{\tau_{\epsilon_i-\epsilon_j}^0=}
{}+\sum_{k\ne
i,j}\biggl(\frac{1}{z_1+z_2}(T_{ik}(z_1)T_{kj}(z_2)+T_{ik}(z_2)T_{kj}(z_1))
-\frac{1}{2z_2}T_{ij}(z_1)T_{kk}(z_2)
\\ \phantom{\tau_{\epsilon_i-\epsilon_j}^0=}
{}-\frac{1}{2z_1}T_{kk}(z_1)T_{ij}(z_2)\biggr)\biggr),
\\
\tau_{\epsilon_i}^{\epsilon_j}=T\biggl(\frac{z_2-z_1}{2z_1z_2(z_1+z_2)}
(T_{ij}(z_1)(T_{ii}(z_2)-T_{jj}(z_2))
+(T_{jj}(z_1)-T_{ii}(z_1))T_{ij}(z_2))
\\ \phantom{\tau_{\epsilon_i}^{\epsilon_j}=}
{}+\sum_{k\ne
i,j}\biggl(\frac{1}{z_1+z_2}\biggl(\frac{1}{z_2}T_{ik}(z_1)T_{kj}(z_2)
+\frac{1}{z_1}T_{ik}(z_2)T_{kj}(z_1)\biggr)
\\ \phantom{\tau_{\epsilon_i}^{\epsilon_j}=}
{}-\frac{1}{2z_1z_2}(T_{ij}(z_1)T_{kk}(z_2)-T_{kk}(z_1)T_{ij}(z_2))\biggr)\biggr).
\end{gather*}

\appendix
\section{Appendix}

In this appendix we want to proof the equations
(\ref{third})--(\ref{Sato33}). For the proof of these equations we
need the following lemma:
\begin{lemma}
\begin{gather*}
\text{\rm Res}_{z}
P(x_i,t,\partial_i)e^{x_iz}(Q(x_i',t',\partial_i')e^{-x_i'z})^T=\sum_j
R_j(x_i,t)S_j(x_i',t')
\end{gather*}
if and only if
\begin{gather*}
\left(P(x_i,t,\partial_i)Q(x_i,t',\partial_i)^*\right)_-=\sum_j
R_j(x_i,t)\partial^{-1}_i S_j(x_i,t')
\end{gather*}
\end{lemma}
This Lemma is a consequence of Lemma  4.1 of \cite{KL}. We rewrite
(\ref{bil4}):
\begin{gather}
\mathop{\rm Res}_z P^{+(0)}(\alpha,\beta,x,t,u,z)
R^{+(0)}(\alpha,\beta,z)Q^+(t,z)e^{x_0z}(P^{-(0)}(\gamma,\delta,y,s,v,-z)\nonumber
\\ \qquad
{}\times R^{-(0)}(\gamma,\delta,-z)Q^-(s,-z) e^{-y_0 z})^T
=\mathop{\rm Res}_z P^{+(1)}(\alpha,\beta,x,t,u,z)
R^{+(1)}(\alpha,\beta,z)\nonumber
\\ \qquad
{}\times Q^+(u,z)e^{x_1z}(P^{-(1)}(\gamma,\delta,y,s,v,-z)
R^{-(1)}(\gamma,\delta,-z)Q^-(v,-z) e^{-y_1 z})^T. \label{bil5}
\end{gather}
First,  applying the Lemma to (\ref{bil5}) gives
\begin{gather*}
(P^{+(0)}(\alpha,\beta,x_0,x_1,t,u,\partial_0)
R^{+(0)}(\alpha,\beta,\partial_0)Q^+(t,\partial_0)(P^{-(0)}(\gamma,\delta,x_0,y_1,s,v,\partial_0)\nonumber
\\ \qquad
{}\times R^{-(0)}(\gamma,\delta,\partial_0)Q^-(s,\partial_0))^*)_-
=\mathop{\rm Res}_z P^{+(1)}(\alpha,\beta,x_0,x_1,t,u,z)
R^{+(1)}(\alpha,\beta,z)Q^+(u,z)\nonumber
\\ \qquad
{}\times
e^{x_1z}\partial_0^{-1}(P^{-(1)}(\gamma,\delta,x_0,y_1,s,v,-z)
R^{-(1)}(\gamma,\delta,-z)Q^-(v,-z) e^{-y_1 z})^T. \label{first}
\end{gather*}
Now putting $s=t$, $u=v$, $x_1=y_1$, we obtain
\begin{gather}
(P^{+(0)}(\alpha,\beta,x_0,x_1,t,u,\partial_0)
P^{-(0)}(\alpha,\beta,x_0,y_1,s,v,\partial_0)^*)_-=0,
\\[1ex]
\Biggl(P^{+(0)}(\alpha,\beta,x_0,x_1,t,u,\partial_0)S(\partial_0)
P^{-(0)}(\alpha-\sum_{j=1}^n \epsilon_j, \beta-\sum_{j=1}^n
\epsilon_j,x_0,y_1,s,v,\partial_0)^*\Biggr)_-=0.\!\!\!
\label{second}
\end{gather}
Since
\begin{gather*}
P^{\pm(0)}(\alpha,\beta,x_0,
x_1,t,u,\partial_0)=I+\sum_{j=1}^\infty
P^{\pm(0)}_j(\alpha,\beta,x_0, x_1,t,u)\partial_0^{-j}\nonumber
\\
P^{\pm(1)}(\alpha,\beta,x_0,x_1,t,u,\partial_1)=\sum_{j=1}^\infty
P^{\pm(1)}_j(\alpha,\beta,x_0,x_1,t,u)\partial_1^{-j},
\label{order}
\end{gather*}
this implies (\ref{third}), one thus also has:
\begin{gather}
P^{-(0)}_1(\alpha,\beta,x_0,x_1,s,v)^T=P^{+(0)}_1(\alpha,\beta,x_0,x_1,s,v).
\label{P+-}
\end{gather}
If we apply the Lemma in the other way we obtain
\begin{gather}
(P^{+(1)}(\alpha,\beta,x_0,x_1,t,u,\partial_1)
R^{+(1)}(\alpha,\beta,\partial_1)Q^+(u,\partial_1)(P^{-(1)}(\gamma,\delta,y_0,x_1,s,v,\partial_1)\nonumber
\\[.5ex] \qquad
{}\times R^{-(1)}(\gamma,\delta,\partial_1)Q^-(v,\partial_1))^*)_-
=\mathop{\rm Res}_z P^{+(0)}(\alpha,\beta,x_0,x_1,t,u,z)
R^{+(0)}(\alpha,\beta,z)\nonumber
\\[.5ex] \qquad
{}\times
Q^+(t,z)e^{x_0z}\partial_1^{-1}(P^{-(0)}(\gamma,\delta,y_0,x_1,s,v,-z)
R^{-(0)}(\gamma,\delta,-z)Q^-(s,-z) e^{-y_0 z})^T. \label{fourth}
\end{gather}
Now taking $s=t$, $u=v$ and $y_0=x_0$, we f\/ind
\begin{gather}
(P^{+(1)}(\alpha,\beta,x_0,x_1,t,u,\partial_1)
R^{+(1)}(\alpha-\gamma,\beta-\delta,\partial_1)
P^{-(1)}(\gamma,\delta,x_0,x_1,t,u,\partial_1)^*)_-\nonumber
\\[.5ex] \qquad
{}\!=\!\mathop{\rm Res}_z\! P^{+(0)}\!(\alpha,\beta,x_0,x_1,t,u,z)
R^{+(0)}\!(\alpha\!-\!\gamma,\beta\!-\!\delta,z)\partial_1^{-1}
P^{-(0)}\!(\gamma,\delta,x_0,x_1,t,u,\!-\!z)^T\!\!. \label{fifth}
\end{gather}
Now choose $\gamma-\alpha=\sum\limits_{i=1}^n \epsilon_i$, then
\begin{gather*}
 R^{+(0)}(\alpha-\gamma,\beta-\delta,z)=
 \sum_{i=1}^n (-)^{i+1}z^{-1}E_{ii}=S(z)^{-1}.
\end{gather*}
Since we have assumed that (\ref{j}) holds for $j=0$
\begin{gather*}
|\delta-\beta|=|\gamma-\alpha|=n.
\end{gather*}
Now choose $\delta-\beta=\sum\limits_{i=1}^n \epsilon_i$, then
(\ref{fourth}) turns into{\samepage
\begin{gather*}
P^{+(1)}(\alpha,\beta,x_0,x_1,t,u,\partial_1)S(\partial_1)\nonumber
\\ \qquad
{}\times P^{-(1)}\Biggl(\alpha+\sum_{i=1}^n
\epsilon_i,\beta+\sum_{i=1}^n
\epsilon_i,x_0,x_1,t,u,\partial_1\Biggr)^*=S(\partial_1)^{-1},
\label{fifth2}
\end{gather*}
thus we obtain (\ref{invP1}).}

Dif\/ferentiate the bilinear identity (\ref{bil5}) to $t_j^{(a)}$,
one obtains
\begin{gather}
\mathop{\rm Res}_z\biggl(\frac{\partial
P^{+(0)}(\alpha,\beta,x,t,u,z)} {\partial t_j^{(a)}}+
P^{+(0)}(\alpha,\beta,x,t,u,z)z^jE_{aa}\biggr)
R^{+(0)}(\alpha,\beta,z)\nonumber
\\ \qquad
{}\times Q^+(t,z)e^{x_0z}(P^{-(0)}(\gamma,\delta,y,s,v,-z)
R^{-(0)}(\gamma,\delta,-z)Q^-(s,-z) e^{-y_0 z})^T\nonumber
\\ \qquad
{}=\mathop{\rm Res}_z\frac{\partial
P^{+(1)}(\alpha,\beta,x,t,u,z)} {\partial t_j^{(a)}}
R^{+(1)}(\alpha,\beta,z)Q^+(u,z)e^{x_1z}\nonumber
\\ \qquad
{}\times(P^{-(1)}(\gamma,\delta,y,s,v,-z)
R^{-(1)}(\gamma,\delta,-z)Q^-(v,-z) e^{-y_1 z})^T. \label{preSato}
\end{gather}
Now using the Lemma  and choosing $\alpha=\gamma$, $\beta=\delta$,
$s=t$, $u=v$ and $x_1=y_1$ this gives the familiar Sato--Wilson
equations (\ref{Sato}). Taking $j=1$  one obtains
\begin{gather*}
\sum_{i=1}^n\frac{\partial P^{+(0)}(\alpha,\beta,x_0,
x_1,t,u,\partial_0)}{\partial t_1^{(i)}}=[\partial_0,
P^{+(0)}(\alpha,\beta,x_0, x_1,t,u,\partial_0)].
\end{gather*}
Return to (\ref{preSato}) and use the Lemma in the opposite way
and choosing, $\gamma-\alpha=\delta-\beta=\sum\limits_{i=1}^n
\epsilon_i$, $s=t$, $u=v$ and $x_0=y_0$ this gives:
\begin{gather*}
\mathop{\rm Res}_z
P^{+(0)}(\alpha,\beta,x,t,u,z)z^jE_{aa}S(z)^{-1}\partial_1^{-1}
P^{-(0)}\Biggl(\alpha+\sum_{i=1}^n
\epsilon_i,\beta+\sum_{i=1}^n\epsilon_i,x,s,v,-z\Biggr)^T
\\ \qquad
{}=\frac{\partial
P^{+(1)}(\alpha,\beta,x,t,u,\partial_1)}{\partial t_j^{(a)}}
S(\partial_1) P^{-(1)}\Biggl(\alpha+\sum_{i=1}^n
\epsilon_i,\beta+\sum_{i=1}^n\epsilon_i,y,s,v,-z\Biggr)^*
\end{gather*}
using (\ref{invP1}) one deduces:
\begin{gather}
\frac{\partial P^{+(1)}(\alpha,\beta,x,t,u,\partial_1)}{\partial
t_j^{(a)}}=\mathop{\rm Res}_z z^{j-1}
P^{+(0)}(\alpha,\beta,x,t,u,z)E_{aa}S(\partial_1)^{-1}\nonumber
\\ \qquad
{}\times P^{-(0)}(\alpha+\sum_{i=1}^n
\epsilon_i,\beta+\sum_{i=1}^n \epsilon_i,x,s,v,-z)^T S(\partial_1)
P^{+(1)}(\alpha,\beta,x,t,u,\partial_1). \label{Sato3}
\end{gather}
Note that we can rewrite (\ref{Sato3}) as (\ref{Sato33}). Since we
have replaced in the wave matrices $t^{(a)}_1$,  by
$t^{(a)}_1+x_0$, (\ref{Sato3}) for $j=1$ implies
\begin{gather}
\frac{\partial P^{+(1)}(\alpha,\beta,x,t,u,\partial_1)}{\partial
x_0}=\mathop{\rm Res}_z
P^{+(0)}(\alpha,\beta,x,t,u,z)S(\partial_1)^{-1}\nonumber
\\ \qquad
{}\times P^{-(0)}(\alpha+\sum_{i=1}^n
\epsilon_i,\beta+\sum_{i=1}^n \epsilon_i,x,s,v,-z)^T S(\partial_1)
P^{+(1)}(\alpha,\beta,x,t,u,\partial_1). \label{SSato3}
\end{gather}

Dif\/ferentiating the bilinear identity to $u_j^{(a)}$ gives that
\begin{gather*}
\mathop{\rm Res}_z\frac{\partial P^{+(0)}(\alpha,\beta,x,t,u,z)}
{\partial u_j^{(a)}}
R^{+(0)}(\alpha,\beta,z)Q^+(t,z)e^{x_0z}\nonumber
\\ \qquad
{}\times(P^{-(0)}(\gamma,\delta,y,s,v,-z)
R^{-(0)}(\gamma,\delta,-z)Q^-(s,-z) e^{-y_0 z})^T\nonumber
\\ \qquad
=\mathop{\rm Res}_z \biggl(\frac{\partial
P^{+(1)}(\alpha,\beta,x,t,u,z)} {\partial
u_j^{(a)}}+P^{+(1)}(\alpha,\beta,x,t,u,z)z^jE_{aa}\biggr)
R^{+(1)}(\alpha,\beta,z)\nonumber
\\ \qquad
{}\times Q^+(u,z)e^{x_1z}(P^{-(1)}(\gamma,\delta,y,s,v,-z)R^{-(1)}
(\gamma,\delta,-z)Q^-(v,-z) e^{-y_1 z})^T. \label{preSato2}
\end{gather*}
Using the Lemma and next choosing, $\alpha=\gamma$,
$\beta=\delta$,  $s=t$, $u=v$ and $x_1=y_1$ we obtain
\begin{gather}
\frac{\partial P^{+(0)}(\alpha,\beta,x,t,u,\partial_0)}{\partial
u_j^{(a)}}P^{+(0)}(\alpha,\beta,x,t,u,\partial_0)^{-1}\nonumber
\\ \qquad
{}=\mathop{\rm Res}_z z^j
P^{+(1)}(\alpha,\beta,x,t,u,z)E_{aa}\partial_0^{-1}
P^{-(1)}(\alpha,\beta,x,t,u,-z)^T. \label{sato2}
\end{gather}
In particular taking $j=1$ one obtains
\begin{gather}
\frac{\partial P^{+(0)}(\alpha,\beta,x,t,u,\partial_0)}{\partial
u_1^{(a)}}P^{+(0)}(\alpha,\beta,x,t,u,\partial_0)^{-1}\nonumber
\\ \qquad
={}-P^{+(1)}_1(\alpha,\beta,x,t,u)E_{aa}\partial_0^{-1}
P^{-(1)}_1(\alpha,\beta,x,t,u,)^T. \label{**}
\end{gather}
Since we have replaced in the wave matrices $u^{(a)}_1$,  by
$u^{(a)}_1+x_1$, (\ref{**})  implies
\begin{gather}
\frac{\partial P^{+(0)}(\alpha,\beta,x,t,u,\partial_0)}{\partial
x_1}\nonumber
\\ \qquad
{}=-P^{+(1)}_1(\alpha,\beta,x,t,u)\partial_0^{-1}
P^{-(1)}_1(\alpha,\beta,x,t,u,)^T
P^{+(0)}(\alpha,\beta,x,t,u,\partial_0) \label{***}
\end{gather}
which is  equivalent to (\ref{sato22}).

Using the Lemma in the other way one deduces for
$\gamma-\alpha=\delta-\beta=\sum\limits_{i=1}^n \epsilon_i$, that
\begin{gather*}
\Biggl(\biggl(\frac{\partial
P^{+(1)}(\alpha,\beta,x,t,u,\partial_1)}{\partial
u_j^{(a)}}+P^{+(1)}(\alpha,\beta,x,t,u,\partial_1)
\partial_1^j E_{aa}\biggr)S(\partial_1)\nonumber
\\ \qquad
{}\times P^{-(1)}\Biggl(\alpha+\sum_{i=1}^n
\epsilon_i,\beta+\sum_{i=1}^n
\epsilon_i,x,t,u,\partial_1\Biggr)^{*}\Biggr)_-=0.
\end{gather*}
Now, using (\ref{invP1}) one deduces the Sato--Wilson equations
(\ref{Sato4}). A special case of (\ref{Sato4}), viz $j=1$ states
that
\begin{gather*}
\sum_{i=1}^n\frac{\partial
P^{+(1)}(\alpha,\beta,x,t,u,\partial_1)}{\partial
u_1^{(i)}}=[\partial_1,P^{+(1)}(\alpha,\beta,x,t,u,\partial_1)].
\label{sixth}
\end{gather*}

\subsection*{Acknowledgements}

This work has been partially supported
 by the European
 Union through the FP6
Marie Curie RTN {\em ENIGMA} (Contract number MRTN-CT-2004-5652)
and the European Science Foundation Program MISGAM.

\pdfbookmark[1]{References}{ref}
\LastPageEnding

\end{document}